\RequirePackage{rotating}
\documentclass[graybox]{svmult}

\usepackage{amssymb,amsmath,amsfonts,eurosym,geometry,ulem,graphicx,caption,color,setspace,sectsty,comment,footmisc,caption,pdflscape,subfigure,array,hyperref}
\usepackage[graphicx]{realboxes}
\usepackage{booktabs}
\usepackage{cleveref}
\usepackage{makeidx} 
\usepackage[numbers]{natbib}
\usepackage{newtxtext}       %
\usepackage{newtxmath} 
\usepackage{rotating}
\newcolumntype{L}[1]{>{\raggedright\let\newline\\arraybackslash\hspace{0pt}}m{#1}}
\newcolumntype{C}[1]{>{\centering\let\newline\\arraybackslash\hspace{0pt}}m{#1}}
\newcolumntype{R}[1]{>{\raggedleft\let\newline\\arraybackslash\hspace{0pt}}m{#1}}

\makeindex

\begin{document}


\title*{Symbiotic Game and Foundation Models for Cyber Deception Operations in Strategic Cyber Warfare}
\titlerunning{Symbiotic Game and Foundation Models for Cyber Deception}
\author{Tao Li and Quanyan Zhu}
\institute{Tao Li (Corresponding author) \at New York University, NY, 11201, \email{tl2636@nyu.edu}
\and Quanyan Zhu \at New York University, NY, 11201 \email{qz494@nyu.edu}
}
%
%
\maketitle

\abstract{
We are currently facing unprecedented cyber warfare with the rapid evolution of tactics, increasing asymmetry of intelligence, and the growing accessibility of hacking tools. In this landscape, cyber deception emerges as a critical component of our defense strategy against increasingly sophisticated attacks. This chapter aims to highlight the pivotal role of game-theoretic models and foundation models (FMs) in analyzing, designing, and implementing cyber deception tactics. Game models (GMs) serve as a foundational framework for modeling diverse adversarial interactions, allowing us to encapsulate both adversarial knowledge and domain-specific insights. Meanwhile, FMs serve as the building blocks for creating tailored machine learning models suited to given applications. By leveraging the synergy between GMs and FMs, we can advance proactive and automated cyber defense mechanisms by not only securing our networks against attacks but also enhancing their resilience against well-planned operations. This chapter discusses the games at the tactical, operational, and strategic levels of warfare, delves into the symbiotic relationship between these methodologies, and explores relevant applications where such a framework can make a substantial impact in cybersecurity. The chapter discusses the promising direction of multi-agent neurosymbolic conjectural learning (MANSCOL), which allows the defender to predict adversarial behaviors, design adaptive defensive deception tactics, and synthesize knowledge for the operational level synthesis and adaptation. FMs serve as pivotal tools across various functions for MANSCOL, including reinforcement learning, knowledge assimilation, formation of conjectures, and contextual representation. This chapter concludes with a discussion of the challenges associated with FMs and their application in the domain of cybersecurity. 
}

\section{Introduction}





Addressing network security challenges is increasingly daunting due to the rise of highly sophisticated attackers, often backed by significant funding and resources from nation-states. These adversaries possess ample resources, including advanced capabilities, dedicated personnel, and deep knowledge, enabling them to pursue their objectives with precision. Traditional defense mechanisms, such as encryption and firewalls, are no longer sufficient in this evolving landscape. In fact, contemporary network security issues have evolved into a form of cyber warfare, wherein adversaries leverage digital technologies to target assets for strategic, political, or military gains. The cyber warfare can be integrated into physical domains.  Adversaries can exploit vulnerabilities in both cyber and physical infrastructure in a coordinated fashion to gain a strategic advantage over their opponents. 

Through offensive operations, employing tactics like the cyber kill chain, adversaries deploy multiple attack vectors to undermine critical systems in a stealthy and persistent manner. In response, defenders are tasked with safeguarding their assets against cyber threats and attacks. The essence of cyber defense lies in outmaneuvering the attacker through innovative strategies and technologies to thwart malicious activities and preserve the integrity of network infrastructure. To this end, recent trends have focused on automated, proactive, and adaptive defenses that are able to reduce cognitive demands, improve operational tempo and mission speeds, gain information advantage, and achieve decision dominance.

Cyber deception emerges as one promising class of defenses, involving the creation of traps, decoys, or false information to mislead attackers and divert their attention from valuable assets or genuine security measures \cite{jajodia2016cyber,al2019autonomous}. The deployment of cyber deception techniques enables network systems to detect, delay, or disrupt attackers' activities while gathering intelligence about their tactics, techniques, and procedures (TTPs). An automated and adaptive cyber deception strategy can proactively predict and respond to attackers' behaviors by creating further deception to deter the attack or engage attackers to gather information. By minimizing the involvement of human operators in the cyber response workflow and intelligently aligning tactical responses with mission objectives, this approach facilitates faster and more mission-aligned decision-making, enhancing cyber resilience for the mission.


\subsection{Role of Game-Theoretic Models}

The interaction between defenders and attackers in cyber deception and the cyber warfare in general is often conceptualized as a strategic game, wherein both parties endeavor to outmaneuver the other to achieve their respective goals. In recent literature, many game models (GMs) serve a descriptive framework, wherein a specific attack model, including strategies, objectives, and information structures, is assumed alongside a defender model \cite{tao_info}. Subsequently, the analysis of the game aims to find the equilibria of the game, which include equilibrium payoffs, strategies, and, in dynamic games, equilibrium dynamics or trajectories.

The overarching aim of this analytical endeavor is twofold. Firstly, it seeks to forecast the long-term outcomes of cyber warfare by evaluating the defender's optimal response to repeated attacks orchestrated by adversaries. Secondly, it endeavors to use equilibrium policies as robust defense mechanisms. These strategies are suitable to serve as default approaches in situations where no real-time information about the attackers is accessible, and hence, there is a need for assumptions about worst-case scenarios to ensure performance-guaranteed strategies. However, in scenarios where information is available, albeit incomplete, through intelligence gathering and online observation, a dynamic adaptation of strategies can be implemented to complement these default policies, thereby enhancing the resilience of defensive measures.

This methodological framework finds applications across diverse domains, including jamming games \cite{zhu2010stochastic,zhu2011eavesdropping,xu2017game,nugraha2019subgame}, authentication games \cite{ge2023gazeta,rass2020cryptographic,ge2022mufaza,gupta2022game,saritacs2019adversarial}, routing games \cite{zhu2012deceptive,zhu2012interference,clark2012deceptive}, intrusion detection \cite{rass2017physical,hu2022evasion,zhu2009dynamic,zhu2010network}, infrastructure protection games \cite{chen2019game,huang2018factored,chen2019dynamic,chen2017dynamic,huang2017large}, and insider threat \cite{huang2021duplicity,casey2016compliance,casey2015compliance,feng2015stealthy,liu2008game}. These endeavors yield valuable insights and establish a fundamental understanding of adversarial interactions at a higher level. Typically, the equilibrium analysis of these models plays a crucial role in informing strategic-level decisions, such as risk assessment, investment decisions, and resource planning within organizations and security contexts. Several factors contribute to the significance of equilibrium. Firstly, the credibility of equilibrium forecasts increases as games are played repeatedly over an extended period. With time, players tend to figure out the structure and the rule of the game, adopt rational strategies, and make more informed decision-making. Secondly, strategic-level decisions do not mandate a high-level precision of the details of implementation and operations and thus can rely on the equilibrium prediction. A rough estimate of the equilibrium outcomes within the correct magnitude suffices for effective high-level decision-making processes.  One area for which game-theoretic frameworks have offered valuable tools is the strategic risk analysis against a variety of adversarial models, each representing contingent goal-driven adversaries \cite{rass2018game,chen2021dynamic,chen2018linear}. This application is particularly vital within cybersecurity contexts, where predicting adversarial behaviors proves challenging, unlike natural events governed by probabilistic distributions. Adversarial TTPs drive these behaviors. Leveraging game theory provides a distinctive framework for cyber risk analysis, with implications spanning domains like cyber insurance \cite{zhang2017bi,zhang2019mathtt,schwartz2014cyber,zhang2021optimal}, cyber-physical system integration \cite{huang2020dynamic,chen2022system,zhu2020cross,huang2020dynamic}, and security investment \cite{cavusoglu2008decision,grossklags2009uncertainty,chen2018security,chen2019interdependent}.

\begin{figure}
\begin{center}
    \includegraphics[scale=0.7]{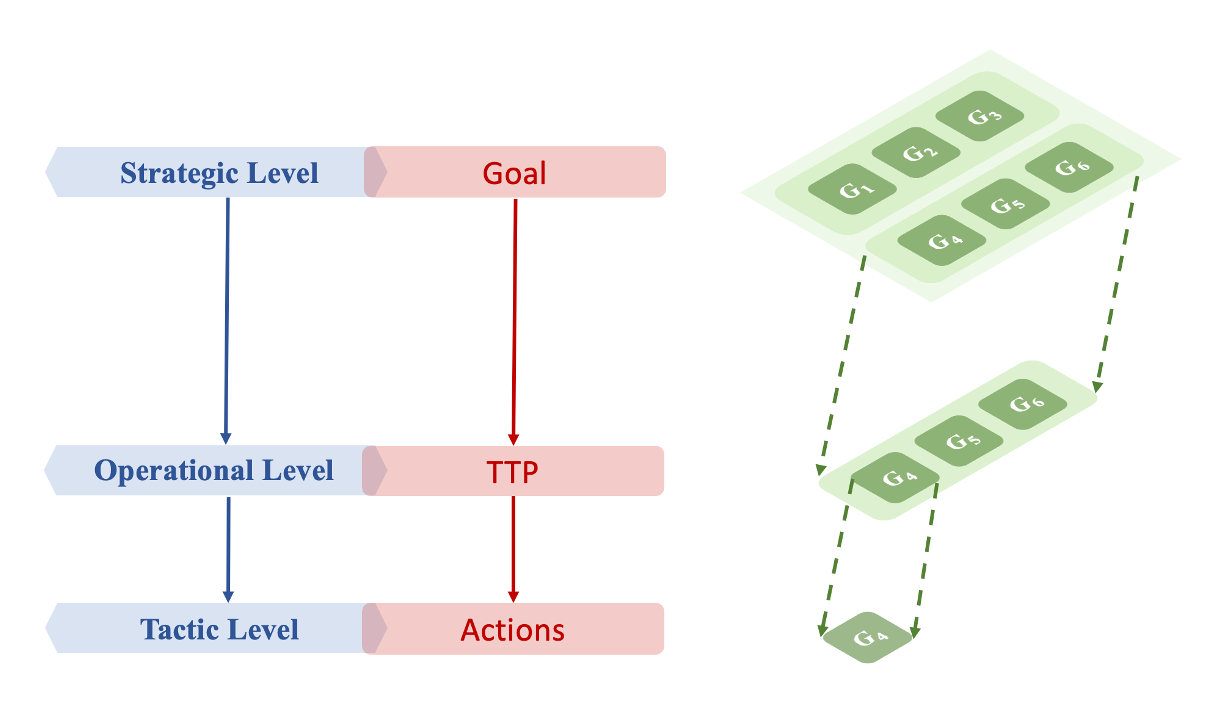}
    \caption{Multi-level game-theoretic frameworks: strategic level, operational level, and tactical level games. Strategic level games are games that describe high-level decision-making, such as resource allocations and investment planning. The goal of strategic level games is to create long-term planning to achieve overarching objectives of the cyber warefare. Tactical-level games involve specific actions and maneuvers that can be implemented to achieve immediate objectives to support the overarching strategy. Examples of tactics in cyber warfare include the configuration of honeypots and the attacker engagement policies. The operational-level games sit between the strategic and tactical levels, focusing on the planning and coordination of a sequence of defense actions. Examples include the planning of a series of cyber defense strategies starting from intelligence gathering to counter lateral movement to achieve strategic level goals.}
\end{center}
\end{figure}\label{game}

\begin{figure}
\begin{center}
    \includegraphics[scale=0.7]{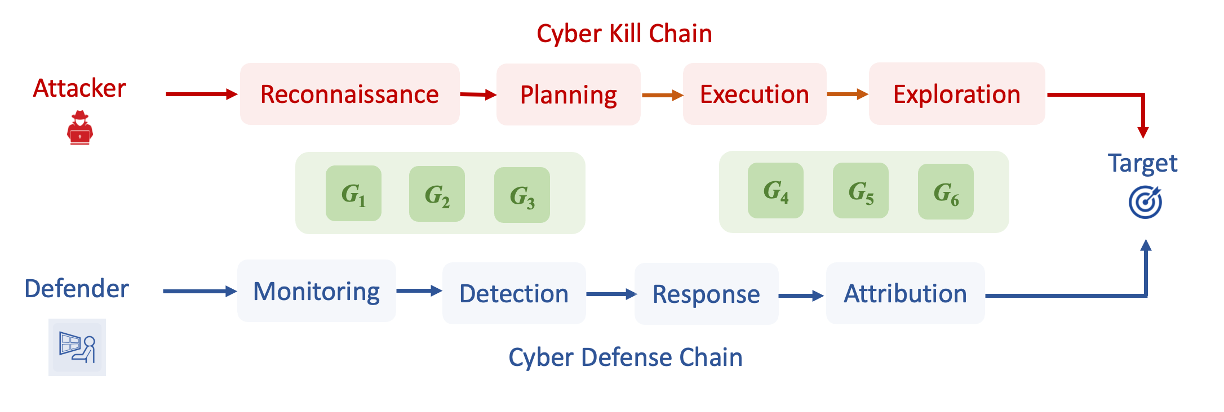}
    \caption{An example of game modeling: An attacker aims to carry out a cyber kill chain to reach the target while a defender aims to deter and thwart this operation. The goal of the attacker is determined through strategic-level reasoning. It can be viewed as an outcome of a high-level game description. Once the goal is set, the cyber kill chain determines the tactics, techniques, and procedures (TTP) to achieve its goal, while the defender determines the defending TTPs. This operation is composed of a sequence of tactic-level games can provide specific techniques and actions. Each tactic level game corresponds to a stage in the operation. The games will yield tactics that determine the outcome at each stage and, eventually, the outcome of the operation. An adaptive operation is often used to reconfigure the operation when the operation fails at certain stages. In this case, the games will need to be redesigned and synthesized to adapt to the uncertainties in the outcomes.}
\end{center}
\end{figure}\label{chain}

However, the lack of precision makes it less applicable to tactical scenarios where details of how to act are imperative in the face of an attack. Obtaining exact models of the game proves challenging, particularly regarding the strategies of attackers and their knowledge. Sometimes, it is also challenging for the defender itself to figure out the knowledge and available strategies. To address this challenge, literature often turns to more intricate modeling approaches, such as Bayesian games \cite{harsanyi1968games,harsanyi1995games} and hypergames \cite{bennett1980hypergames,wang1989solution}, to accommodate uncertainties. In particular, in the domain of cyber deception, where information asymmetry is prevalent, Bayesian games have emerged as an appropriate class for describing cyber deception scenarios. These scenarios involve various tactics such as honeypot configuration \cite{la2016game,la2016deceptive,boumkheld2019honeypot}, attack engagement \cite{huang2019dynamic,huang2020dynamic,caballero2024behavioral}, and information design \cite{zhang2023stochastic,zhang2022forward,zhang2021equilibrium}, among others.

While the Bayesian approach provides new insights into games of incomplete information by formally modeling imperfect observations and uncertainties about game variables, it shifts the challenges toward uncertainty quantification of underlying unknowns, introducing new complexities to the modeling process. As uncertainties multiply, especially when dealing with numerous uncertain parameters or structures, the model's complexity escalates. For instance, when uncertain parameters are continuous variables, quantifying uncertainty in random variables becomes a formidable task.

Uncertainty quantification often requires a substantial amount of data or repeated interactions between attackers and defenders within the same environment to establish reliable distributions. However, obtaining such data poses challenges, as the game is not time-invariant, and the rounds of interactions between attackers and defenders are limited, which further constrains the observables obtained from these interactions.

Moreover, while this approach is reasonable for handling structured uncertainties (i.e., known unknowns), it encounters difficulties with unstructured uncertainties (e.g., epistemic uncertainties). Furthermore, even with a well-defined model, computing equilibrium concepts like Bayesian equilibrium proves challenging \cite{tao23pot} and often lacks predictive power, especially in dynamic tactical environments characterized by nonstationary behaviors and conditions.

It is imperative to go beyond the current approach.  Recently, nonstationary reinforcement learning and nonequilibrium solutions have emerged as prominent ways to tackle this challenge. Nonstationary reinforcement learning aims to tackle the nonstationary nature inherent in learning environments. In security games, an agent's game can undergo changes due to the dynamics of the knowledge and behaviors of other players. Additionally, the joint actions taken by the players of the game can lead to shifts in the game environment. It is one type of internalized nonstationarity, which is caused by the players themselves. Externalized nonstationarity refers to changes caused by exogenous factors; for instance, in cyber warfare, the disclosure of vulnerabilities to the public can alter the game dynamics between defenders and attackers due to the change in the information structure.

Learning in a nonstationary environment poses challenges as strategies adapted to earlier games must be re-evaluated for new environments \cite{tao_info}. Unlike in stationary environments, where learning strategies often converge to relevant equilibrium concepts \cite{tao22confluence}, nonstationary games cannot be characterized by equilibrium, and there is a need for a reevaluation of reinforcement learning goals. Learning strategies do not need to aim to converge to an equilibrium outcome \cite{shutian23erm}; rather, an equilibrium outcome can be viewed as a result of learning behaviors \cite{pan-tao22noneq}. This perspective can lead to the nonequilibrium solutions that redefine outcomes for nonstationary games \cite{li2024automated}. Nonequilibrium solutions are relevant even in stationary games where interactions are limited, meaning players cannot reach equilibrium within short timeframes.

\subsection{Role of Foundation Models}
Such approaches offer promising avenues to navigate the complexities of dynamic tactical environments where traditional equilibrium concepts may prove insufficient. The advent of foundation models (FMs) introduces a new dimension of possibility to this endeavor. These models enable the encoding of game descriptions through more sophisticated frameworks, offering enhanced precision in capturing detailed aspects of the game. For instance, FMs can encapsulate the knowledge structure of attackers, non-Markov dynamics (e.g., time series data), and diverse attacker types. This granularity holds the potential to provide tactical-level solutions, thereby facilitating improved decision-making processes.

Moreover, FMs facilitate the modeling of flexible learning and reasoning processes of varying styles. Unlike current approaches that are often pre-specified in terms of learning methods and objectives, this adaptability allows for the creation of novel solution concepts capable of describing outcomes within limited interactions, nonstationary environments, and non-Markovian behaviors. Such models are particularly suitable for generating practical, real-time tactics that can achieve decision dominance.

Furthermore, FMs enable not only descriptive modeling but also prescriptive and predictive features. Prescriptive analytics directly yield end-to-end intelligence that informs tactics, bypassing the separate process of modeling the entire game, defining and finding the associated solutions, and creating decision-dominant tactics. This end-to-end approach facilitates the translation of information directly into actionable and adaptive tactics.

Another pivotal aspect of FMs lies in their predictive analytics capability, which plays an important role in warfare scenarios. The ability to anticipate and promptly respond to emerging threats can be a decisive factor. Predictive analytics serves several critical functions in tactical reasoning. Firstly, it enables defenders to forecast the dynamic environment, including the anticipated behaviors of adversaries and the external factors influencing environmental shifts. This foresight enables proactive measures to mitigate risks. Secondly, predictive analytics integrates into the learning and adaptation of strategies, enabling the formulation of look-ahead strategies. These strategies analyze subsequent subgames and anticipate potential counter-strategies from attackers, thereby creating decision-dominant tactics in cyberwarfare. The incorporation of predictive analytics into FMs enhances their adaptability and effectiveness in navigating dynamic and adversarial environments.

\subsection{Cyber Deception and Related Game-Theoretic and Foundation Models}
 Cyber deception is an overt operation. Defenders need to design mechanisms to outsmart attackers, leading them to fall into honeypots or unwittingly disclose information beneficial to the defender's objectives. The application of FMs to cyber deception games is an ideal synergy.  FMs can contribute to this goal by improving analytical capabilities crucial for strategic reasoning. A significant challenge in cyber deception lies in the inherent uncertainties. The environment is rife with unknowns, including imprecise knowledge of attacker attributes. It remains unclear whether attackers are aware of the deception and may deploy counterdeception tactics. Additionally, the behaviors of non-attackers, such as normal users, pose another layer of uncertainty. While deceiving normal users is not the intent of cyber deception, their errors can inadvertently diminish its efficacy. FMs offer invaluable tools for addressing these challenges, facilitating hierarchical and predictive reasoning to enhance decision-making processes. For instance,  models can anticipate attacker responses to deception and utilize cognitive hierarchies to optimize decisions. Predictive reasoning enables the anticipation of both attacker actions and user behaviors in subsequent steps, thereby preparing defenders to preempt attacks and mitigate user errors effectively.

The essence of cyber deception resonates strongly with game-theoretic models. Cyber deception has essential characteristics that align closely with game-theoretic attributes, including the information asymmetry between the players, multi-stage and multi-phase interactions, the presence of aleatoric and epistemic uncertainties, together with the bounded rational human behaviors.  In \cite{pawlick2019game}, a taxonomy of cyber deception games is developed to connect each deception scenario with a fundamental class of game-theoretic models. The taxonomy leads to a library of game building blocks that can be used to synthesize multi-round multi-scale dynamic games that capture the cyber kill and defense chains for strategic, operational, and tactic level decision-making. For example, honeypot engagement has been captured by a class of stochastic games with one-sided information. Honeypot deployment problems have been captured by network design games. Intrusion evasion games have been captured by a class of principal-agent detection games. These building blocks enable a divide-and-conquer and modular approach to design cyber defense. These modular games can be fused together with FMs to create function modules that are capable of representing games in the security scenario to provide strengthened learning and computation power. This enables the development of tactical-level applications and the enhancement of cyber deception techniques. 

With the aid of game theoretic frameworks, which capture the intricate dynamics of cyber deception, the informed use of FMs becomes possible. Game-theoretic frameworks facilitate the training and adaptation of FMs to suit contextual applications with symbolic representation of the attacker-defender interactions.  Integrating game-theoretic models with FMs can play useful roles across different phases of the cyber deception process. During the initial design phase, FMs serve as invaluable tools for training and preparation. Subsequently, they enable reinforcement learning, enabling adaptability and refinement as the cyber deception strategy evolves. In the post-implementation phase, FMs support adaptive tuning, ensuring ongoing optimization and effectiveness of cyber deception techniques.

There are several ways to integrate the both. One approach is neurosymbolic learning, where hybrid AI algorithms combine symbolic reasoning with data-driven learning, resulting in robust and trustworthy systems. Symbolic reasoning using game-theoretic models offers the capacity to incorporate sophisticated abstractions grounded in system and game theories and associated formalisms. Supported by advanced tools and methods, neurosymbolic learning enables integrated analysis and assurance, leveraging formal specification and verification technologies. These capabilities can be instrumental in strengthening cyber deception mechanisms. Likewise, meta-reinforcement learning represents another avenue for integrating FMs with game theory. This approach involves developing models capable of learning from a diverse array of game-theoretic security contexts rather than being restricted to a singular context. Meta-reinforcement learning builds on the trained models to create adaptive and self-improving cyber deception systems.

Meta-reinforcement learning and neurosymbolic learning can be integrated to form an iterative process where game-theoretic domain knowledge evolves based on observations, while defense policies adapt using reinforcement learning methodologies. This process intertwines data-driven updates with domain-specific game-theoretic primitives to result in a hybrid neural and symbolic representation. This process can be empowered by FMs, and it will converge towards an optimal combined hybrid neural and symbolic representation.

Large language models (LLMs) stand out as pivotal FMs well known for their proficiency in handling textual and heterogeneous data. Their unparalleled capacity to comprehend extensive textual datasets makes them indispensable for natural language processing tasks, enabling the processing of text, time-series data, and predictive analytics. In cybersecurity, they empower the augmentation of network security through human-like language comprehension and predictive capabilities. Within the domain of cyber deception, LLMs can generate honey files such as counterfeit documents and credentials to entice potential attackers. Moreover, LLMs excel in processing heterogeneous datasets encompassing network traffic, system logs, and user behaviors to detect anomalies, infer potential attacker behaviors, and predict malicious activities. When encountering evolving threats,  LLMs adeptly can automate cyber deception policies to mitigate risks by putting together multiple sources of information, including new vulnerability disclosure, defender's service requirements, and the current system state.

This chapter aims to introduce the symbiotic relationship between game-theoretic models and FMs, focusing on their implications in cybersecurity and cyber deception domains. Figure \ref{guardware} illustrates this relationship. FMs serve as the building blocks for AI applications and research, providing developers and researchers with a solid platform to pioneer and innovate in artificial intelligence and machine learning. Security games stand as building blocks in cybersecurity applications, offering contextual and reasoned frameworks that encapsulate the interactions between attackers and defenders across diverse scenarios. The synergy of these models promises a transformative approach to designing \textbf{guardware} that provides autonomous and proactive security measures. In particular, the integration lays the groundwork for designing cyber deception mechanisms that were used to be regarded as cumbersome, limited, or challenging. By leveraging these foundational designs, we can forge a new paradigm of intelligent security mechanisms and win cyber warfare. 

\begin{figure}
\begin{center}
    \includegraphics[scale=0.7]{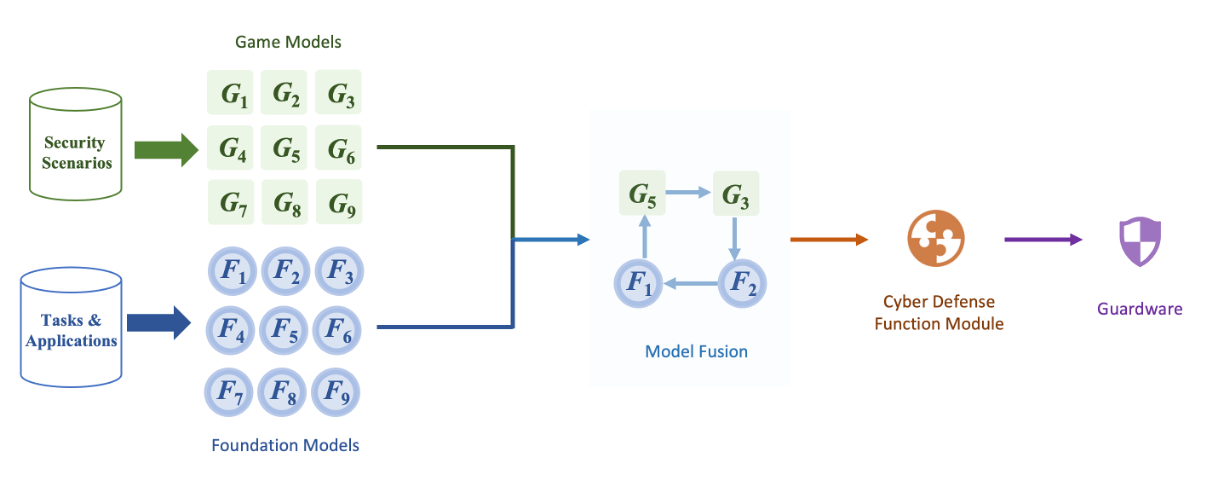}
    \caption{Game-theoretic models and FMs are fused together to create function modules for cyber defense, which will be built into the guardware. Game-theoretic models are representations of security scenarios while FMs are tailored for different tasks and applications. The guardware is composed of multiple function modules that are enabled different games and FMs. Each function module requires a different architecture to synthesize game and FMs.}
\end{center}
\end{figure}\label{guardware}

\subsection{Organization of the Chapter}
This chapter is organized as follows. Section 2 provides the background on cyber deception and network security games. It discusses the game-theoretic approach to cyber adversarial modeling and the associated analytical, design, and learning approaches. Section 3 provides a background on FMs and their roles in AI, learning, and their relevance in cybersecurity.
Section 4 presents the synergy of game-theoretic models and FMs in the context of cyber deception. Section 5 presents several challenges and future directions that need to be addressed to develop transformative security technologies. Chapter 6 concludes this chapter.

\section{Cyber Deception and Network Security Games}


Security games capture the intricate interactions between attackers and defenders in various contexts. These scenarios span from cybersecurity applications, involving the interaction between intruders and network system administrators, to critical national security applications where safeguarding pivotal assets, such as airports and infrastructures, is paramount against potential terrorist threats. The formalization of security games frequently relies on game-theoretic languages, specifying essential elements like payoffs, players, and action sets.

Cyber deception stands out due to its unique tactics and objectives, which involve the strategic deployment of decoys, honeypots, and false information to manipulate attackers' perceptions of the network and influence their actions. The primary aim of deception is twofold: to deter attacks from exploiting unknown vulnerabilities within our systems and to gain valuable intelligence from attackers' behaviors. Game-theoretic models for cyber deception often revolve around information structures. For instance, asymmetric information games elucidate the information asymmetry between two players. Signaling games serve as a fundamental framework for understanding two-player deception scenarios, where one player observes the true state of the world while the other player can only observe messages or actions. They have been applied to study detection evasion \cite{hu2022evasion}, honeypots \cite{pawlick2018modeling}, and insider threats \cite{casey2016compliance}.

Stochastic games with one-sided information, e.g., in \cite{horak2017manipulating,li2024conjectural,anwar2020game}, extend this framework into dynamic settings, where one player has complete observation of the state space while the other player navigates with incomplete information, forming beliefs over the state space. Information acquisition games facilitate the trade-off between the cost of gathering information and the quality of decision-making based on such information. Recent literature has been investigating the design of observation kernels \cite{liu2023information,zhang2020deceptive} and optimal timing for information acquisition \cite{huang2019continuous,huang2021pursuit,huang2021defending}. 

\subsection{Multi-Level and Multi-Scale Game Models}
Decision-making in cybersecurity is multi-scale. We define three levels of modeling, which require distinct levels of granularity. They are, namely, strategic level, operational level, and tactic level. Cybersecurity problems are growing nowadays into a cyber warfare. An attacker plans a kill chain that involves multiple stages of attacks to achieve his planned goals. At the same time, a defender aims to deter and thwart the operation of the attacker to protect the target. The strategic level involves long-term planning and decision-making aimed at achieving overarching goals and objectives. The strategies are concerned with defining the overall goals, allocating resources, and determining the target to attack or protect. The tactical level involves specific actions and maneuvers implemented to achieve immediate objectives and support the overarching strategy. The tactics include the techniques, procedures, and actions that can be taken or implemented to carry out a specific task. Examples include the tactics to engage an attacker in a honeypot and the attacker's tactics to do lateral movement to get to the target. The operational level sits between the strategic and tactical levels, focusing on the planning and coordination of multi-stage tactics to achieve the specified mission from the strategic level. The operation involves planning and replanning the multiple steps of tactics to achieve the goal of compromising a targeted machine. 

As illustrated in Fig. \ref{game}, the GMs involved at the three levels are multi-scale. At the strategic level, the game is a high-level, coarse-grain description of the adversarial interactions between a defender and an attacker. For example, Blotto-type games, e.g., \cite{roberson2006colonel,hart2008discrete,golman2009general} have been classically used to allocate a limited amount of resources to overpower adversaries in cyber warfare by outnumbering their resources. FlipIt games and their variants have also been recently studied in \cite{van2013flipit,laszka2014flipthem,bowers2012defending} to strategically analyze the interplay between attackers and defenders, both striving to assert control over a resource for the maximum duration while minimizing overall move costs. Through strategic reasoning, defenders devise counterstrategies aimed at controlling resources at opportune moments and minimizing downtime and potential compromise.
Network games are another class of strategic-level security games. Networks are used to capture the interdependencies and connectivity among the resources. The goal of the defender is to protect or defend essential network assets and minimize the impact of the attacks.  For example, the framework proposed by \cite{8673619} introduces a network design paradigm allowing network designers to fortify links to establish redundant communication pathways between nodes to mitigate potential attacks through resource investment. By proactively considering strategic cyber threats, the framework offers methods to characterize and compute optimal strategies for securing a network within predefined budget constraints.  In addition to protective measures, resilient strategies for network recovery following attacks are also essential in network defense. A two-player, three-stage game framework proposed in \cite{chen2017dynamic} has aimed to capture both protection and recovery phases. The network designer's objective is to maintain network connectivity both before and after an attack, while the adversary seeks to disrupt the network by compromising specific links. This strategic game framework facilitates analysis of trade-offs not only among nodes and links within the network but also among resources allocated for pre-event protection and post-event recovery.

The game at the operational level decomposes the objective of attacking and defending a selected asset into multiple stages of operations. In accordance with the cyber kill chain, attackers typically execute reconnaissance stages before planning attacks, which involve footprinting, social engineering, and intelligence gathering. Subsequently, after planning, the attack mission is executed through the exploitation of vulnerabilities, privilege escalation, and lateral movement to reach the targeted asset. For this operational sequence, the reconnaissance stage can be modeled as a series of games that delineate the process of information gathering through passive or proactive means. The attacker's planning following reconnaissance can also be modeled through a sequence of games, encompassing multiple stages such as vulnerability exploitation, lateral movement, command and control, and exfiltration. Planning can adapt if the attack fails at any stage, necessitating reevaluation and adaptation of acquired observations and gathered information.

An example of the operation-level games is described in \cite{zhu2018multi}. A multi-phase and multi-stage game is used to capture the attack vectors of advanced persistent threats. Each phase and each stage has distinct characteristics and requires a different form of defense strategy. A dynamic game framework can piece together games at each phase and each stage to capture the whole attack-and-defense operation holistically from the entry point to the target. A more sophisticated dynamic game that incorporates incomplete information and cyber deception has been presented in \cite{huang2020dynamic} and \cite{huang2019dynamic}. It depicts long-term dyadic interaction between a stealthy attacker and a proactive defender through a multi-stage game of incomplete information. Each player holds private information unknown to the other, and strategic actions are guided by beliefs formed through multi-stage observation and learning processes.  

The game at the tactic level is represented by the short-term local interactions between an attacker and a defender. Each stage of the operation involves a distinct type of dyadic interactions. For example, in the social engineering game at the reconnaissance stage, the attacker can exploit the cognitive vulnerabilities of the defenders to gain initial access. The attacker often has an information advantage when it aims to deceive the defender. At the stage of lateral movement, the defender can guide the attacker to a honeypot so that the attack will be detected and revealed. This defensive deception provides the defender information advantage at the stage of the game. Hence, throughout the operation, the game can take different forms depending on the structure of information, interactions, and maneuverability. The survey in \cite{pawlick2019game} explores game theory frameworks designed to model defensive deception strategies in cybersecurity and privacy domains. It introduces a taxonomy comprising six distinct types of deception: perturbation, moving target defense, obfuscation, mixing, honey-x, and attacker engagement. These classifications are characterized by their information structures, agents involved, actions taken, and duration, all of which align closely with game theory concepts.  This taxonomy establishes a structured framework for comprehending the diverse strategies of defensive deception.  The survey outlined in \cite{manshaei2013game} offers a curated selection of studies that demonstrate the application of game theory in addressing various security and privacy challenges within computer networks and mobile applications. In each domain, we identify security problems, players, and GMs, including security of the physical and MAC layers, security of self-organizing networks, intrusion detection systems, anonymity and privacy, economics of network security, and cryptography. 
 The games that describe the local interaction can be viewed as a building block to synthesize an operational level game, and hence informing the creation of the strategic level games. 

The multi-level interdependencies among the three levels of the games are illustrated in Fig. \ref{game}. The required granularity of the GMs differs across levels. The strategic-level models are long-term and require a high-level description of potential operation consequences once resources are utilized. The outcome of the strategic-level Blotto game can be predicted by an operational-level game simulator. The tactical-level models require higher precision, as precision is crucial for determining the exact actions to be taken at the focused stage. The tactical-level games are driven by the high-level objectives specified at the strategic level. Hence, inherent interdependencies exist among the three layers of the games. The tactical ones inform the consequences of the strategic ones, while the strategic ones specify the objectives, constraints, and structures of the tactical games. The operational ones reside in the middle layer, serving as an intermediary that interconnects both ends. 

It can be observed that cyber warfare games are multi-scale and multi-resolutional. Zooming in from the strategic level to the tactical level is a top-down process that specifies elements of the tactical games. Conversely, zooming out from the tactical level to the strategic level is a bottom-up process. The outcomes of individual tactical games and their compositions lead to the prediction of strategic level interactions and inform strategic level decisions. The zooming-in and zooming-out process can be adaptive. When strategic-level decisions change, the tactical level needs to adapt to the change in an agile manner. If tactics fail due to uncertainties, unexpected events, or errors, resulting in operational-level task failures, there is a need to resiliently adjust strategic-level decisions. This adjustment allows for the design of and adaptation to new missions to recover from the failure and ensure defense.

Accompanied by the ability to zoom in and out, new solution concepts beyond equilibrium are necessary to predict game outcomes across various levels and resolutions. These concepts must be compatible with uncertainties, limited observations, and the non-stationary nature of interactions. Hence, nonequilibrium solutions are essential to address these features, aiming to create a reasonable prediction even though the outcome of the game can be stochastic with limited rounds of interactions. Moreover, solutions for tactical-level games must be consistent with those at the strategic level, despite differences in scales or levels of detail. This consistency represents another dimension of consistency, in addition to conjectural and incentive consistencies (e.g., associated with the solution concepts Berk-Nash equilibrium \cite{esponda2016berk}, Nash equilibrium \cite{nash1953two}), leading to a holonic solution where localized tactical games consistent with global strategic ones.

The new equilibrium concepts play a crucial role in assessing the risk of the overall mission. Mission risk is inherently multi-scale; risks at the tactical level can impact those at the strategic level. Moreover, planning decisions' risks at the strategic level can influence the selection of tactical level strategies. High-level risks can impose constraints on low-level strategies and guide their decision-making process. Conversely, low-level risks can propagate and manifest as high-level risks. Computational tools associated with solution concepts at each level must be integrated to create a cohesive tool capable of automated, coordinated, zoom-in, and zoom-out risk analysis.

Learning and adaptation are crucial aspects of the solution framework. Rather than viewing equilibrium solutions as defining the learning process, we must reconsider and understand that learning processes shape game outcomes. Learning serves as a natural descriptor of gameplay, alongside strategic and extensive form game descriptions. This description can result in equilibrium and non-equilibrium outcomes depending on the time horizon examined. 

The learning description of the game facilitates the development of adaptation schemes for games at multiple levels, as adaptation can be directly and descriptively designed rather than being considered a derivative of equilibrium analysis. Just as solution concepts associated with multilevel learning need to be consistent, adaptation across levels must also exhibit consistency. Firstly, adaptation must exhibit top-down causality, where adaptation at the tactical level may be triggered by uncertainties or unexpected events at its own level and by adaptation at the strategic level. Secondly, adaptation must be coordinated and stable. Multilevel adaptation requires coordination to avoid scenarios where adaptation at one level causes more failures or leads to cyclic or unstable outcomes.

\subsection{FM-Enabled GMs}
 The perspective described above delineates a new research direction aimed at developing programmable and mosaic game-theoretic models along with a range of versatile methods to advance the design of decision-dominant defense tactics and strategies to prevail in cyber warfare. The initial phase involves establishing a library of tactical games as foundational components. These tactic games serve as fundamental building blocks that encapsulate the typical adversarial situations encountered by defenders. They encode the defender's knowledge regarding feasible actions and potential attack responses using various representations, including strategic form games, extensive form games, and learning form games.

Strategic and extensive form games have traditionally been discussed in textbooks. For instance, a simple rock-paper-scissors game can be depicted using a matrix, which is a form of strategic representation. Meanwhile, a stochastic dynamic game with multiple stages of interactions can be represented using a tree structure that illustrates the multi-round interactions among players, their observations, and the uncertainties inherent in decision-making. The learning form game represents a novel approach, specifying how agents respond to and learn from acquired information and knowledge. It eliminates the necessity to utilize utility functions to define players' objectives or incentives. Instead, players' objectives are implicitly conveyed through dynamic learning processes wherein they adapt their strategies toward long-term goals. These goals are expressed through sequences of strategy updates, potentially leading to either equilibrium or non-equilibrium outcomes. The learning representation facilitates the translation of observed player behaviors into learning patterns, which may be nonstationary. The uncertainty quantification of behaviors using learning represents a promising avenue for bridging theory and practice. Furthermore, the GMs offer immediate adaptability, an inherent property of learning dynamics. Moreover, the way in which players learn can also adapt to contextual or environmental changes.

A tactical game is designed to be composable with others. This composability can result in a larger game, leading to a multi-phase and multi-stage operational game that describes a more complex cyber operation \cite{zhu2018multi}. Composability can take on different forms: sequential, expansive, or reinforcing. Sequential composability forms a sequential game where the scenario of one game is followed by that of another. It allows the defender to prepare for forthcoming interactions and make informed defense decisions before the scenario changes, potentially shifting the advantage from the defender to the attacker. Expansive composability involves aggregating two games into one larger game with a more complex structure of action spaces, dynamics, or information structures. It can be viewed as a parallel composition where players engage in two scenarios simultaneously. This enables the defender to confront an attacker with augmented capabilities or multiple attackers coordinating to achieve their goals. Reinforcing composability is a type of feedback composability where the outcome of one game affects the outcome of the other and vice versa. It captures gameplay in which one cyber scenario influences another and eventually returns to itself. When the feedback has a negative impact, the defender faces a vicious cycle where failure leads to further losses in the same scenario.

 The canonical forms of composability enable the game framework to be both mosaic and adaptive. By combining two composable games, defenders can effectively analyze and strategize against attackers. Moreover, as contexts or strategic objectives evolve, these games can be reconfigured or recomposed to align with new missions. This flexibility is particularly crucial during operational setbacks, where swift adaptation is vital for strategic resilience. Such adaptability introduces a novel capability for defenders in cyber warfare, and it can be further strengthened by advances in FMs that facilitate online learning and knowledge representation.

The development of tactical games and their ability to compose, reason, and learn is not limited to online response and interactions with adversaries.  Another advantage lies in the development of game-theoretic digital twins capable of simulating hypothetical scenarios and predicting outcomes previously encountered \cite{tao2024dima}. The capabilities of game-theoretic digital twins can inform strategic-level decision-making, allowing comprehensive planning by evaluating risks associated with both familiar and unfamiliar scenarios and the reasoning of an optimal mission-driven strategy. Furthermore, at the tactical level, the game-theoretic digital twin can also be used to create conjectures or predictions of the adversarial behaviors, allowing the defender to gain a preparatory advantage over the attacker and improving the tactical decision dominance.


Learning and adaptation occur at multiple levels, forming several adaptation loops. Strategic-level adaptation must coordinate with associated adaptations at the tactical level. A rapid top-down translation between strategic-level mission goals and tactical-level tactics is necessary to enable agile responses to changes in the continually evolving threat landscape in cyberspace. This translation necessitates different levels of knowledge representation. At the strategic level, knowledge and information are encoded as semantic objectives, rules, constraints, and feasible sets \cite{peng21local,yin24pirl}. At the tactical level, knowledge and information are represented by acquired spatio-temporal observations encoded in numerical values, time-series data, and binary outcomes. Neurosymbolic learning, which integrates symbolic reasoning with data-driven learning, is essential to facilitate top-down and bottom-up coordination of responses to heterogeneous knowledge and information generated at different levels and from different sources.

This translation operates in both directions. Apart from the top-down impact of knowledge on tactical decisions, new knowledge generated at the tactical level can significantly influence decisions at operational and strategic levels. The propagation of knowledge, both bottom-up and top-down, requires careful study. It demands symbolic learning mechanisms operating at multiple scales to effectively disseminate insights and optimize decision-making processes. To support these investigations, there is a need to establish new solution concepts along the way. Non-equilibrium and learning-based solution concepts are essential for describing the outcomes of multi-level games. These solutions must demonstrate consistency across levels, while the learning process should be causal, coordinated, and stable.



FMs allow a more powerful way to achieve the above.  They enhance the learning capabilities of each game building block, particularly in tactical games where higher resolution payoffs, information, and dynamics are required. FMs excel in understanding information from diverse sources, including textual, semantic, and numerical data, facilitating the fusion of information and enhancing game resolution through FM-enabled analytics.

Moreover, FMs facilitate the synthesis of GMs. FMs can automate the integration and synthesis of operational games by selecting and fusing relevant sets of tactical games. Additionally, they provide a concise and high-level depiction of strategic-level games for decision-making purposes, akin to how LLMs summarize information and grasp its essence.

FMs can also play a role in the coordination of learning and adaptation across the layers. Reinforcement learning enabled by the FMs can provide augmented capabilities for adaptation and learning, leveraging the multi-source and multi-modal time-series information. To achieve this, new architectures involving FMs and GMs are needed to achieve cross-level coordination and adaptation.

Fig. \ref{architecture} illustrates the fundamental architectures of the FMs with GMs. A GM can be supported by an FM to identify the game through the learning of the incentives, information structure, and dynamics of the players and provide a more accurate representation of the interactions between the players. This relationship has been depicted in Fig. \ref{architecture}(a). Two GMs can also be integrated to create a meta-game through an appropriate FM. This process creates operational-level games from the game building blocks at the tactical level. This architecture is illustrated in Fig. \ref{architecture}(b). Illustrated in Fig. \ref{architecture}(c), the sequential adaptation and coordination between two GMs, each supported by their FMs, are essential to create tactical level resiliency when encountering uncertainties and time-varying environments. The coordination between the two stages at the tactical level improves the efficacy of the operation. When the mission changes, the operation needs to adapts too. The adaptation of an operation meta-game can be enabled by an FM, which not only reacts to the changes in the mission and outcomes of the tactics but also coordinates the learning at the tactical levels. This cross-level architecture is depicted in Fig. \ref{architecture}(d).

\begin{figure}
\begin{center}
    \includegraphics[scale=0.7]{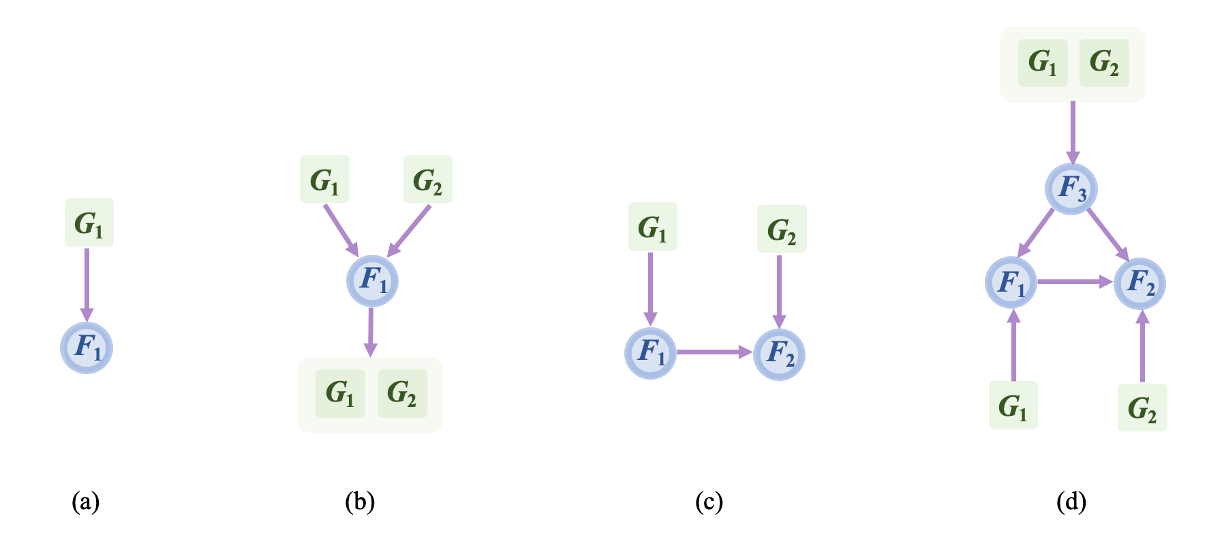}
    \caption{Examples of Architectures of Symbiotic FMs and GMs: (a) FM $F_1$ enriches GM $G_1$ by improving the precision of the GM and augmenting its learning capabilities; (b) Tactical games $G_1$ and $G_2$ are fused together to achieve an operation game through the FM; (c) The adaptation of two sequential tactic games $G_1$ and $G_2$ is enabled by FM $F_1$ and $G_2$. They are coordinated sequentially;  (d) The operational level game learning $F_3$ coordinates the learning of $F_1$ and $F_2$. }
\end{center}
\end{figure}\label{architecture}



\section{Foundation Models and Large Language Models and the Synergetic Roles in Cybersecurity}

FMs refer to machine learning models that are trained on broad data, generally using self-supervision at scale, such that they can be adapted to a wide range of downstream tasks. FMs differ from existing machine learning (ML) models in two aspects: scale and scope. Scale: FMs generally feature large-scale neural network models with an astronomical number of parameters to be determined by self-supervised training over comparable data. For example, GPT-3, an FM in natural language processing, has 175 billion parameters and is trained on 45TB of text data.  Scope: FM, as a generalist, aims to handle multi-modal data and perform a wide range of tasks without being explicitly trained to do so. It is observed that FMs possess few-show generalizability: a handful of demonstrations from new tasks are sufficient for FMs to adapt.   In contrast, ML models are specialists, targeting specific tasks and acting on certain types of data, such as image classification. They typically suffer from poor generalization when transferred to tasks unseen in the training stage. In summary, FMs are large-scale generative models with highly resource-intensive pretraining, capable of learning general features and patterns from diverse data sources.   Examples of FMs include large language models (LLM) for natural language processing \cite{openai23}, DALL-E for images\cite{ramesh21dalle}, MusicGen for music \cite{copet2024simple}, and RT-2 for robotic control \cite{zitkovich23rt2}.  

\begin{figure}[!h]
    \centering
    \includegraphics[width=1\textwidth]{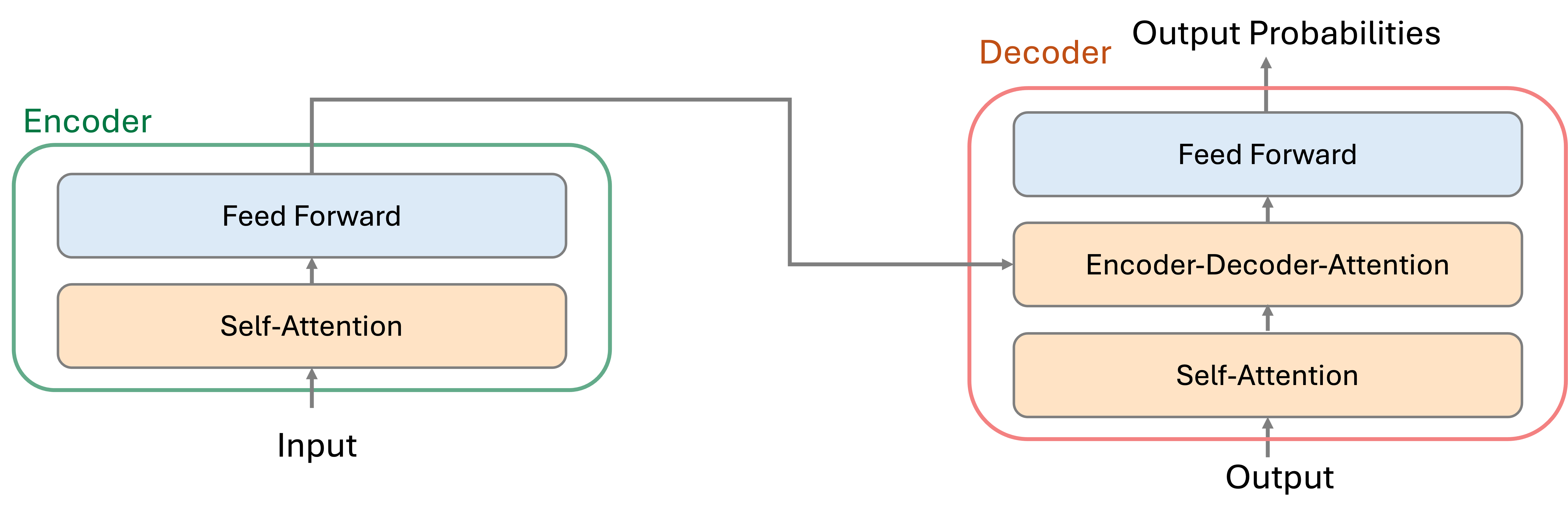}
    \caption{The encoder-decoder structure of the Transformer architecture, adapted from \cite{vaswani17transformer}. Each input datapoint, after embedding and positional encoding, is fed to the attention module in the encoder part (the right half), which extracts temporal correlation (attention scores) across the input datapoint sequence. The attention scores are then passed to the decoder (the left half) to generate an output sequence auto-regressively, together with additional attention within the output sequence. The attention mechanism is instrumental in descriptive, predictive, and perspective analytics in cybersecurity.}
    \label{fig:transformer}
\end{figure}
The few-shot generalization ability of FMs stems from the attention mechanism in the transformer architecture \cite{vaswani17transformer} shown in \Cref{fig:transformer}, which effectively captures long-range temporal dependencies among inputs. As the building block of FMs, the transformer consists of an encode (on the left half) and a decoder (on the right half). The task of the encoder is to map an input sequence from multi-modal high-dimensional data sources to a sequence of latent representations in a unified vector space, which is then fed to the decoder. Upon receiving the encoder's output, the decoder generates a new output based on its previous outputs, producing an output sequence in an auto-regressive manner. Mathematically, denote the input sequence by ${\mathbf{x}}=\{x_1,x_2, \ldots, x_n\}$, which, for example, can be a sentence to be translated or a sequence of sensor readings for robot controls.  The transformer is a parameterized auto-regressive probabilistic model $\mathcal{P}_{\theta}$ that consumes the input $\mathbf{x}$ and generates an output sequence $\mathbf{y}=\{y_1,y_2,\ldots, y_m\}$ using previously generated outputs, i.e., $y_t\sim \mathcal{P}_\theta(\cdot|y_{t-1},\ldots, y_1; \mathbf{x})$.  

Each input datapoint $x_i$ to the transformer, which we will call a token, is initially embedded in vectors of a certain embedding dimension with additional positional encoding so that each position in the input sequence acquires a unique representation. Each input embedding generates a query, key, and value vector of dimensions $d_k$, $d_k$, and $d_v$. Vectors of the same type are stacked column-wise to produce three matrices $Q \in \mathbb{R}^{n \times d_k}$, $K \in \mathbb{R}^{n \times d_k}$, and $V \in \mathbb{R}^{n \times d_v}$, with $n$ being the length of the input sequence. The attention score is then calculated with the formula
\begin{equation*}
\text{Attention}(Q, K, V) = \text{softmax}\left(\frac{QK^\intercal}{\sqrt{d_k}}\right) V.
\end{equation*}
The matrix $QK^\intercal$ is divided by $\sqrt{d_k}$ to prevent the vanishing gradient problem when applying the soft-max row-wise \cite{vaswani17transformer}. The entries of the resulting attention matrix represent the correlation among input tokens. Using the translation example, each entry captures how the word to be translated relates to others in the sentence. Both the encoder and decoder rely on the attention mechanism for correlation extraction in the original sentence $\mathbf{x}$ and the translation $\mathbf{y}$, respectively. Yet, their usage of attention bears a subtle difference. The upper off-diagonal triangle part of the resulting matrix  $\text{softmax}({QK^\intercal}/{\sqrt{d_k}})\in \mathbb{R}^{l \times l}$ is masked with 0s in the decoder part. This causal mask prevents future tokens from influencing the prediction of the current output and is the defining feature of a causal transformer, distinctive from other variations. 

To allow the attention mechanism to extract correlation from different representations, FMs often utilize multi-head attention where queries, keys, and values are linearly projected multiple times (one projection corresponds to one head attention) with different learned projections to $d_k$, $d_k$, and $d_v$ dimensions, respectively. The attention mechanism reviewed above acts on the projection in parallel, producing $d_v$-dimensional values to be concatenated into a single final value, as shown in \Cref{fig:transformer}. The positional encoding and generation of $Q$, $K$, $V$ matrices, together with feed-forward networks and linear layers in the transformer, are trainable components. The parameter size soars up and easily attains the billion level for recent FMs \cite{openai23}, as the embedding dimension, context length, and number of heads increase.  

The attention mechanism is the cornerstone of FM's generalization when performing downstream tasks. The key observation is that the correlation is invariant across various sequential data, and the attention score acquired in pre-training lends itself to a variety of game-theoretic security analytics across various dimensions presented below.

\subsection{Descriptive Analytics}
FMs demonstrate their use by harnessing historical data to construct a formal description of game scenarios, capturing the dynamics, incentives, and players involved in the interactions. For instance, drawing upon past experiences, these models can create a matrix game that encapsulates the competitive interplay between an inspector, employing randomized examinations to identify attackers, and the agile attacker, adept at evading scrutiny. This descriptive analytics framework captures the essence of the evolving interactions within security games.

Another emerging game-theoretic security paradigm brought by this descriptive analytics is the natural game representation. Unlike the matrix and other standard game representations in the cybersecurity context, the natural representation features an end-to-end treatment of the actual network systems through FMs and, in particular, large language models (LLM) as the prevalent data in cybersecurity are of text type. LLMs are capable of summarizing evidence-based insights from lengthy cyber threat intelligence (CTI) texts and deriving semantic knowledge from CTI \cite{cybert21}, which serves as the game state. Thanks to the attention mechanism, LLMs can discover the correlations between vulnerabilities and attack patterns, mapping Common Vulnerabilities and
Exposures (CVE) to CommonWeaknesses Enumeration (CWE) \cite{cve-cwe} and linking CVEs to MITRE ATT \& CK techniques \cite{cve2attck}. Such LLM-based mapping naturally specifies the attacker strategy space without undergoing mathematical modeling. The end-to-end security pipeline centered around LLMs and FMs is projected to take an increasing share in cybersecurity task automation.    

\subsection{Predictive Analytics}
The predictive capabilities of FMs extend to anticipating the forthcoming games and deciphering the strategies likely to be employed by adversaries. Given the nonstationary nature inherent in many applications, predictive analytics serves as a strategic tool for gaining an intelligence advantage. By inferring patterns and trends from past interactions, FMs facilitate proactive decision-making, ensuring defenders are well-prepared for the next move in the dynamic landscape of security games.

FMs' predictive power proves to be instrumental in predictive reinforcement learning in the cybersecurity context \cite{li2024automated}, where the policies are updated online according to the defender's forecast of the future consequences and anticipation of the attacker's reactions. Unlike offline or batch reinforcement learning \cite{tao2020causality}, this predictive learning paradigm emphasizes the learning agent's online adaptability to a subjective forecast of future environment evolution and opponent \cite{li2024conjectural}

Two primary challenges arise from the predictive learning paradigm, to which FMs offer a unified data-driven approach. The first is to adapt the policy on the fly to the forecast with lightweight computation. Existing efforts utilize the multistep lookahead idea whereby the agent predicts multiple rounds of possible interactions (trajectories) in the future \cite{tao23cola,li2024automated}. A policy improvement is then carried out by seeking the optimal policy maximizing the cumulative utilities within the lookahead horizon. As pointed out in \cite{tao23cola}, such a lookahead optimization is more involved than vanilla policy optimization. The adapted policy shall only target a set of plausible forecasts, i.e., those close to the actual trajectory evolution, while discarding all other candidates with marginal probabilities. Consequently, the lookahead adaptation boils down to constrained stochastic optimization, requiring sophisticated machinery for efficient computation.     

FMs, as a generative model, naturally fit the role of the predictor in generating future trajectories. Unlike the Monte Carlo simulation that is commonly used in existing works \cite{li2024automated, li2024conjectural}, FMs can extract the temporal correlation among past interactions via attention scores and apply them to generate new forecasts without explicitly modeling the environment, leading to high-resolution model-free predictions that are indistinguishable from actual trajectories \cite{janner2021offline, tao24picol}. The high-accuracy predictions pave the way for online planning algorithms, such as Monte Carlo tree search and rollout methods considered in \cite{tao23cola} and \cite{li2024automated}.        

The second challenge pertains to the agent's subjective perceptions of the environment and the opponent, based on which it makes forecasts. The agent needs to calibrate its subjective conjecture on the environment dynamics \cite{li2024automated} and the opponent's strategy \cite{li2024conjectural} using the online information feedback, e.g., observations. Such a calibration process, performed also online, aims to ensure consistency between the learning agent's subjective perceptions and the objective multi-agent decision-making \cite{li2024conjectural}: what one observes does not contradict what one believes. To ensure consistency, existing works resort to Bayesian learning approaches, inspired by Bayesian parametric statistics \cite{tao23cola, li2024automated, li2024conjectural}. The proposed Bayesian learning begins with a parametric representation of external uncertainties regarding the environment and the opponent: a set of parameterized models is available to the learning agent, each of which represents a possible system dynamics and opponent strategy. Starting with a prior distribution over these models, the agent continuously updates the posterior upon receiving information feedback, which proves to concentrate asymptotically on the models closest to the actual environment dynamics and opponent response \cite{li2024automated, li2024conjectural}.

FMs can revolutionize the calibration process in predictive reinforcement learning. To begin with, it is natural to treat different FMs as parametric models in Bayesian learning, with each pre-trained on different data sources capturing various temporal correlations in system trajectories and opponent sequential actions. Beyond such a straightforward extension, FMs can directly cater to the calibration process in a model-free manner, as investigated in \cite{tao23sce}. Without building parametric models beforehand, FMs cut straight to the opponent's action sequence predictions (forecasts), which are then fed to FMs again to generate the ego agent's actions conditional on its predictions. The calibration takes place implicitly in the transformer's auto-regressive operation: the updated forecasts depend on the previous predictions and the information feedback. Recall that the transformer corresponds to a probabilistic model $y_t\sim \mathcal{P}_\theta(\cdot|y_{t-1}, \ldots, y_1;\mathbf{x})$. In the context of predictive reinforcement learning, $\mathbf{x}$ denotes the past information feedback and the output $y_{t}:=(\widehat{s}_{t:t+K}, \widehat{a}_{t:t+K}, a_t)$ includes the the agent's forecast of future $K$-step system evolution $\widehat{s}_{t:t+K}$,  the opponent's action sequence $\widehat{a}_{t:t+K}$, and the agent's best response $a_t$ against predicted opponent's moves. 

More than a forecaster that generates future trajectories, FMs also assume the roles of the actor (policy improvement) and critic (policy evaluation) in reinforcement learning, leading to an integrated forecaster-actor-critic (FAC) pipeline \cite{li2024conjectural} embodied by the transformer. In contrast to model-based predictive learning \cite{li2024automated, li2024conjectural}, FM-powered FAC opens promising avenues to model-free predictive reinforcement learning, exploring a rich class of opponent/system modeling in a data-driven fashion, alleviating the model misspecification in Bayesian learning \cite{li2024automated}.                  

\subsection{Prescriptive Analytics} 
The prescriptive analytics in cybersecurity aims to provide targeted recommendations for defense strategies, with the overarching objective of minimizing the impact of potential attacks. FMs serve as a mechanism to generate bespoke recommendations and solutions. These actionable insights become invaluable resources for policymakers and network administrators. By leveraging FMs, prescriptive analytics directly enhances the security intelligence available to end-users and defenders. This strategic augmentation ensures a proactive and informed approach to defense, equipping stakeholders with the foresight and tailored solutions needed to mitigate risks effectively.  

The use of prescriptive analytics, together with descriptive and predictive analytics, extends to the design of sophisticated learning algorithms. One emerging paradigm that can benefit from FMs' prescriptive capabilities is the non-equilibrium learning \cite{pan-tao22noneq,pan-tao23delay, pan2024variational} and its associated learning defense. Unlike conventional game-theoretic learning, non-equilibrium learning shifts the focus from long-term equilibrium-seeking to transient strategic interaction characterization. Such a paradigm shift is particularly relevant to cybersecurity, where the defender must acquire an advantageous position within a short window before the attack cyber kill chain materializes \cite{li23dd-ztd}. As delineated in \cite{pan-tao22noneq}, the three pillars of non-equilibrium learning are the target set (e.g., advantageous positions), the measurement function, and the time window. A learning process is said to achieve non-equilibrium if the measurement of the learning dynamics falls in the target set within the time window. Compared with equilibrium-focused multi-agent learning, non-equilibrium learning emphasizes finite-time approachability to a set of desired outcomes \cite{Tao_blackwell}. Thanks to the measurement function, the approachability evaluation in non-equilibrium learning is much more flexible than the utility-driven one in standard game-theoretic learning, which takes into account broader defense objectives than simply utility-maximization, such as privacy and transparency preserving \cite{pawlick2019game,tao23pot}.

The introduction of FMs to the cybersecurity domain enables an offline pre-trained online adaptable prescription for non-equilibrium-based cyber defense. Beginning with the offline pre-training, FMs first digest historical security data, such as CWE, CVE, and system log files, through self-supervised representation learning. The pre-trained FMs provide high-confidence situational awareness of the network systems and adversarial behaviors, which further translates to proper measurement functions producing accurate cyber risk assessment and corresponding desirable outcomes for non-equilibrium learning design. 

Naturally, designing learning dynamics from scratch may take a prolonged period, and the defense may completely miss the window of cyber advantage when the defender takes charges before the attack cycle completes. A more viable and effective defense prescription to achieve non-equilibrium is to consider the meta-learning prescription, whereby the FMs first extract common defense strategies or defense response patterns in the offline stage by consuming diverse security data at scale. When deployed online, a handful of online observations on the network system suffices for fast defense policy adaptation, as FMs are decent few-shot learners due to the attention mechanism \cite{openai23}. The offline pre-trained defense is referred to as the meta-policy, pertaining to a wide variety of security scenarios. As demonstrated in \cite{tao23ztd, pan-tao23meta-sg}, such as meta-learning prescription is scenario-agnostic in the sense that FMs need not examine the exact system configuration. The online adaptation is purely data-driven with marginal computational overhead, leading to effective non-equilibrium defense under system uncertainties.


As we delve into these advanced methodologies, it becomes imperative to reassess traditional equilibrium concepts that underpin our understanding of game outcomes, as motivated by the non-equilibrium concept \cite{pan-tao22noneq}. The conventional objective has been to compute the equilibrium of the game, where the player's uncertainties regarding the environment and opponents are internalized as incomplete information \cite{solan_game}. However, the advent of FMs, where agents harness data in diverse ways to formulate strategies, challenges the adequacy of solutions that internalize all uncertainties. Instead, practicality dictates the exploration of solutions corresponding to models that externalize uncertainties.

Consider predictive reinforcement learning with parametric models representing internalized system uncertainties. In a case where no external uncertainties are introduced, the firm conclusion is that the learning cannot exceed the optimality under the given uncertainty modeling. Yet, such a claim only holds for closed systems. In real-world applications, many uncertainties are not modeled or are impossible to model precisely, which necessitates an open system model with external uncertainties.  

As such, the pursuit of equilibrium defense is no longer legitimate in cybersecurity, due to the presence of externalities. In this context, the need for novel solution concepts arises. As discussed in the preceding subsection, the notion of consistency \cite{li2024conjectural}, inspired by the Berk-Nash \cite{esponda2016berk} and self-confirming equilibrium \cite{self-confirming}, offers a promising avenue for describing game outcomes in models that explicitly externalize uncertainties, achieving superb performance in reinforcement learning \cite{tao23sce} and cybersecurity benchmarks \cite{li2024automated}. In addition to consistency, another relevant notion is decision dominance \cite{li23dd-ztd}, where the dominant party refers to one who can extract information from data and learn about external uncertainties. Whoever completes the information acquisition and analysis cycle faster gains an information advantage over the opponent on the external uncertainties. Unlike classical equilibrium defense resting on the steady state of strategic interactions, decision dominance defense focuses on establishing an information advantage within a transient window and acting decisively before losing the initiative. This shift in perspective acknowledges that not all uncertainties need to be explicitly modeled within the game framework.


\subsection{Foundation models for Mechanism Design for Security Games}

The design of security games is another domain where FMs can contribute to achieving given security objectives with underlying specifications. The formulation of design problems is facilitated through hyper-FMs, wherein FMs serve as fundamental building blocks for the design process. An illustrative example involves the use of FMs by security game designers to determine optimal game parameters, influencing the course of actions undertaken by the involved agents. This process needs to use additional FMs to anticipate the actions and behaviors of the agents.

Many components of interactions are amenable to design or control, including incentives (player payoffs) \cite{huang2022advert}, game dynamics (course direction shaped by actions), and information disclosed to players during gameplay \cite{yating-tao23information, zhang2023stochastic}. The design process itself can take either an offline or online approach. Offline design involves training FMs to ascertain optimal game configurations, while online design employs a reinforcement learning process that dynamically adapts to player actions. Both approaches leverage hyper-combinations of analytics through FMs or the design of hyper-FMs to achieve their objectives.



\section{Cyber Deception Game and Foundation Models}

 There are several quintessential domains in which FMs can play a significant role in cyber deception. Cyber deception operations involve numerous tactical components, including information gathering, detection and response, configuration and deployment of honeypots, and engagement with attacks. In this section, we specifically discuss the connection between FMs and GMs in the use of defensive cyber deception for attack engagement and defensive response. Once an attacker is in the network, the goal of defensive cyber deception is to thwart and mislead the attacker who are attempting to breach or infiltrate the network or systems by deploying techniques such as fake credentials and honeypots to deceive and divert attackers' attention away from valuable assets and towards simulated or less critical targets. Many game-theoretic approaches have been proposed to describe various scenarios, \cite{pawlick2019game,kamhoua2021game}. For instance, a signaling game framework has been used to capture the asymmetric information between the two players \cite{pawlick2018modeling}. In \cite{horak2017manipulating}, a one-sided information stochastic game is used to examine the effects of deception on the attacker's belief. The study has explored the sequential nature of attacks and investigates how an attacker's beliefs evolve, influencing their actions. It has demonstrated strategies for defenders to manipulate attacker beliefs effectively, hindering attackers from achieving their objectives and minimizing network damage. 

\subsection{Challenges in Cyber Deception}
One key challenge inherent in cyber deception lies in understanding attacker behaviors through proactive engagements. It is crucial to infer the incentives and motivations driving attackers by leveraging heterogeneous sources of data, including traffic data, log data, and event data. By doing so, defenders can gain insight into TTPs used by the attackers and their objectives. It enables the defender to prescribe the most effective responses to emerging threats.    

The following \Cref{tab1} is an example of the parsing of PCAP data that captures the scenario when an attacker responds to the patching of network systems and aims for data exfiltration.  Their behavior may change as they adapt to the evolving security measures.  The attacker initially attempts to exploit known vulnerabilities in the target system. However, as the target system patches these vulnerabilities, the attacker's exploit attempts become unsuccessful. The attacker responds by continuing to probe the target for other potential vulnerabilities and, upon identifying a successful attack vector, exfiltrates data to an external server.
This PCAP table illustrates an attacker's adaptive behavior in response to network system patching, highlighting their persistence and ongoing efforts to exploit vulnerabilities and compromise the target environment. 
\begin{sidewaystable}
\begin{tabular}{|c l l l l l l|} 
 \hline
Frame & Action & Source & Destination & Protocol & Flags \\ 
 \hline\hline
1 & \texttt{SYN packet sent} & \texttt{Attacker\_MAC} & \texttt{Target\_MAC} & TCP & SYN   \\ 
 \hline
2 & \texttt{SYN-ACK} response & \texttt{Target\_MAC} & \texttt{Attacker\_MAC} & TCP & SYN  \\ 
 \hline
3 & \texttt{Exploitation attempt} & \texttt{Attacker\_MAC} & \texttt{Target\_MAC} & TCP & PSH ACK  \\ 
 \hline
4 & \texttt{Exploit attempt unsuccessful} & \texttt{Target\_MAC} & \texttt{Attacker\_MAC} & TCP & RST \\ 
 \hline
5 & \texttt{SYN packet sent} & \texttt{Attacker\_MAC} & \texttt{Target\_MAC} & TCP & SYN \\  \hline
6 & \texttt{SYN-ACK response} & \texttt{Target\_MAC} & \texttt{Attacker\_MAC} & TCP & SYN ACK \\  \hline
7 & \texttt{Data exfiltration attempt} & \texttt{Attacker\_MAC} & \texttt{External\_Server\_MAC} & TCP & PSH ACK \\   \hline
8 &  \texttt{Data exfiltration successful} & \texttt{External\_Server\_MAC} & \texttt{Attacker\_MAC} & TCP& ACK \\ 
 \hline
\end{tabular}
\caption{The presented PCAP table shows the adaptive nature of an attacker encountering patched network systems. Initially, the attacker seeks to exploit known vulnerabilities within the target system. However, as the target's vulnerabilities are patched, the attacker's exploit attempts are thwarted. Undeterred, the attacker adapts by persistently probing the target for alternative vulnerabilities. Upon discovering a successful attack vector, the attacker shifts tactics, opting to exfiltrate data to an external server.}
\label{tab1}
\end{sidewaystable}

\begin{sidewaystable}
    \begin{tabular}{|c l l l l l l|} 
 \hline
Frame & Action & Source & Destination & Protocol & Flags \\ 
 \hline\hline
1 & \texttt{Reconnaissance by hesitant attacker} & \texttt{Attacker\_IP} & \texttt{Network\_Router\_IP} & TCP & SYN   \\ 
 \hline
2 & \texttt{SYN packet forwarded to honeypot} response & \texttt{Attacker\_IP} & \texttt{Honeypot\_IP} & TCP & SYN  \\ 
 \hline
3 & \texttt{Honeypot responds with SYN-ACK} & \texttt{Honeypot\_IP} & \texttt{Attacker\_IP} & TCP & SYN ACK  \\ 
 \hline
4 & \texttt{Attacker sends ACK to establish connection} & \texttt{Attacker\_IP} & \texttt{Honeypot\_IP} & TCP & ACK \\ 
 \hline
5 & \texttt{Honeypot acknowledges connection} & \texttt{Honeypot\_IP} & \texttt{Attacker\_IP} & TCP & ACK \\  \hline
6 & \texttt{Attacker probes honeypot cautiously} & \texttt{Attacker\_IP} & \texttt{Honeypot\_IP} & TCP & PSH ACK \\  \hline
7 & \texttt{Honeypot responds to attacker probing} & \texttt{Honeypot\_IP} & \texttt{Attacker\_IP} & TCP & PSH ACK \\   \hline
8 &  \texttt{Attacker engages further with honeypot} & \texttt{Attacker\_IP} & \texttt{Honeypot\_IP} & TCP& PSH ACK \\ 
 \hline
\end{tabular}
\caption{The table illustrates a strategic interaction between a cautious attacker and a honeypot deployed by the defender. The attacker initially conducts a cautious scan of the network for potential targets, exhibiting hesitant behavior. The honeypot strategically responds to the attacker's SYN packet with a SYN-ACK, mimicking a legitimate service to attract the attacker. The attacker acknowledges the honeypot's SYN-ACK by sending an ACK packet to establish a connection. Subsequently, the attacker proceeds cautiously, sending probing packets to the honeypot to gather information without revealing their true intentions. The honeypot responds to the attacker's probing, maintaining the illusion of a legitimate service and engaging in interactive communication.}
\label{tab2}
\end{sidewaystable}

Another example of the attacker's behavior is illustrated using the following PCAP data in \Cref{tab2}. The attacker interacts with honeypots in a discretionary manner. The attacker initially exhibits hesitant behavior, cautiously interacting with the honeypot to gather information and assess the situation. The honeypot, strategically placed by the defender, maintains the illusion of a legitimate service, effectively attracting and engaging the attacker without raising suspicion. This strategic approach allows the defender to gather valuable intelligence about the attacker's behavior and intentions.

A comprehensive understanding of the attacker's response is pivotal for defenders to formulate effective strategies and implement appropriate defensive measures. Creating a robust strategy involves not only understanding the attacker's immediate actions but also anticipating their potential moves. It requires the ability to agilely analyze the attacker's behavior and determine the most suitable course of action to mitigate the impact of the intrusion.

The defense strategy entails several crucial steps. Firstly, defenders must meticulously map out the attacker's TTPs. This involves gaining insights into the attacker's modus operandi, including their preferred attack vectors, tools, and methodologies. By understanding the attacker's playbook, defenders can proactively identify vulnerabilities and develop countermeasures to thwart potential threats.
Moreover, defenders must be adept at estimating the possible trajectories of the attacker's actions. This entails forecasting the potential avenues that the attacker may pursue during the course of the intrusion. By anticipating various scenarios and their associated risks, defenders can better prepare for contingencies and respond effectively to emerging threats.
Central to the success of defense strategy is the ability to act quickly and decisively. Defenders must be equipped to make rapid and well-informed decisions in response to evolving threats. This requires leveraging advanced technologies and methodologies to automate the decision-making process and streamline response efforts.

FMs play a pivotal role in facilitating the learning process and empowering defenders to navigate the complexities of cyber threats effectively. By harnessing FM's capabilities, defenders can gain valuable insights into the attacker's knowledge and intentions, enabling them to map out the cyber landscape and identify potential vulnerabilities. Various types of FMs serve distinct purposes in the intrusion response process. For instance, LLMs excel at parsing and comprehending the nuanced nuances of the attacker's intent, enabling defenders to decipher malicious communications and identify potential threats.
Similarly, Decision Transformers \cite{janner2021offline,chen21dt,tao23sce} leverage sophisticated algorithms and reinforcement learning techniques to predict attacker behaviors and make optimal decisions in real time. By analyzing patterns and trends in attacker activity, Decision Transformers empower defenders to anticipate threats and implement proactive defense measures.

Apart from supporting the ability of the defender to use his knowledge to predict the attacker's paths, FMs can also be used to allow the defender to synthesize and acquire knowledge over time. Understanding the attacker's prior behavior and potential scenarios is crucial for the defender to quickly update its security policies. Proficient knowledge enables the defender to adapt promptly and avoid extensive trial-and-error efforts. This knowledge is typically encoded using symbolic representations of rules, laws, or constraints. For instance, rules may be expressed through first-order or temporal logic statements, while laws can be encapsulated in rule-of-thumb equations, offering approximate responses. Constraints capture feasible directions for updates, which can be weighted to prioritize specific directions. Acquiring knowledge presents a challenge. One immediate method involves knowledge sharing, where insights on unknown attacks can be gleaned from those who have encountered similar incidents. In \cite{fung2011smurfen,zhu2011game,zhu2012guidex}, mechanisms for knowledge sharing among intrusion detection systems are explored, introducing an incentive-compatible approach to encourage data sharing and expedite responses to zero-day attacks. Alternatively, knowledge can be synthesized from individual experiences and data. FMs play a pivotal role here, encoding experiences and data into symbolic representations conducive to formal reasoning, computation, and optimization of security policies. Recent works in \cite{west2022symbolic,wu23llm-symbolic} provide promising approaches of using FMs to synthesize knowledge from datasets. 


Another challenging area of cyber deception is the configuration and utilization of honeypots in engaging attackers to acquire information and knowledge. Honeypot games have been used to formalize the deployment and configuration decisions by taking into account the adversarial interactions and the engagement goals. For instance, in \cite{huang2019adaptive}, a reinforcement learning algorithm is proposed to map out an attacker's behavior. The engagement has to interact with the attacker, making sure that the attacker does not exist in the honeynetwork.  By observing and analyzing attacker interactions with honeypots, defenders gain valuable insights into emerging threats and attacker behaviors. There is a tradeoff between learning the attacker's behaviors and removing the attacker directly. The knowledge of attackers can be useful to further protect ourselves from zero-day attacks in the future. However, removing the attackers directly can lead to fewer risks.

Despite extensive research on honeypot games in both academic and practical domains, a noticeable gap persists between the game-theoretic solutions proposed in the literature and the practical requirements for effective defense mechanisms. A significant challenge lies in the necessity for decision dominance in decision-making processes. This means the defender must make faster, better-informed decisions to safeguard the network effectively. To achieve this, it becomes imperative to leverage both offline knowledge databases about the attacker and online observations of attacker behavior to predict their capabilities, moves, and incentives. FMs are capable of generating different attack scenarios. By leveraging these scenarios, the defender can better prepare for potential outcomes and develop proactive deception tactics by simulating them and making decisions based on the simulated outcomes. By doing so, defensive decisions can account for various contingent scenarios, minimizing risks, and maintaining decision dominance.  

 

\subsection{A Neurosymbolic Learning Approach to Cyber Deception}
\begin{figure}
\begin{center}
    \includegraphics[scale=0.7]{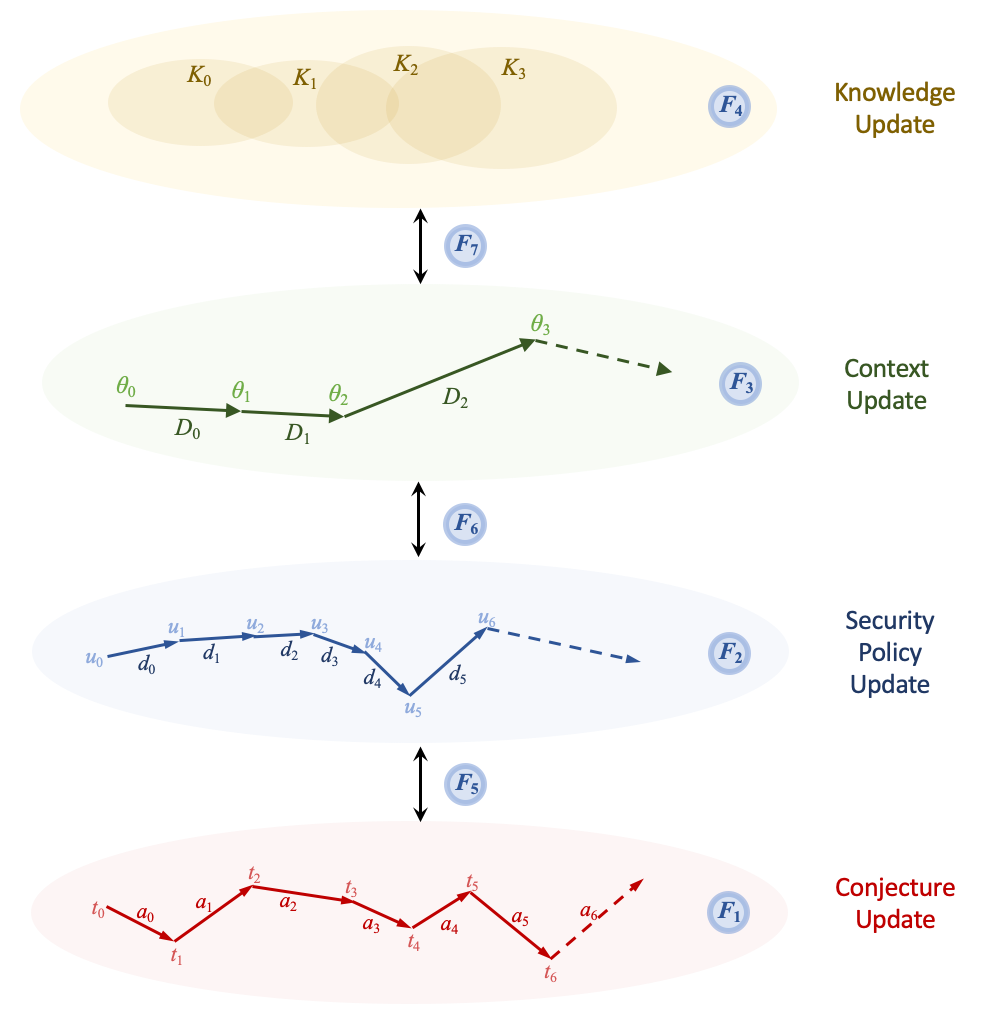}
    \caption{Multi-agent neurosymbolic learning enabled by GMs and FMs: The defender conjectures the attacker's behavior and policies $a$ and finds the optimal security policy update $d$ at each round of tactical interaction between the two players. The defender updates the context parameter $\theta$, which captures the contextual information, including the operation level information and the environmental information. Each context is associated with a knowledge set, which includes a set of rules of the game, including the system constraints on the security policy updates, the admissible set of conjectures, and learning patterns to update the policy updates. FMs are used for reinforcement learning, knowledge acquisition, conjectural formation, and representation of contextual information. In addition, FMs are used to coordinate the building blocks.}
    \label{nslearning}
\end{center}

\end{figure}

In this context, the framework of neurosymbolic learning emerges as a promising approach. This framework integrates multiple FMs to form a cohesive system capable of learning from heterogeneous data sources and making informed decisions regarding attacker engagement strategies. By incorporating neurosymbolic learning techniques, defenders can enhance their ability to adapt proactively to evolving threats and effectively defend against sophisticated cyberattacks. Illustrated in Figure \ref{nslearning}, the defender conjectures the attacker's tactics $t_i$ and policies $a_i$ at each step $i\in T$. Based on the conjectures, the defender designs an optimal policy update $d_i$ to adapt its defense action $u_i$ at each step $i\in T$. The multi-stage rounds of interactions correspond to a sequence of tactical-level games. As the environment and the operational level objective change, the tactical-level games need to adapt. This change is captured through a contextual parameter $\theta_i$, which learns from the heterogeneous sources of data and adapts itself to identify the exact context so that the right set of tactical games can be picked to design the security update policies. The update of $\theta_i$ is at an operational level, which has a different time scale from the tactical level. Let $\bar{T}$ be the set of time steps, which leads to a sequence of updates at a slower frequency in comparison with the one associated with $T$. The contextual update policy is denoted by $D_i, i\in \bar{T}$. Its learning is driven by information from online monitoring of the attack behaviors and network changes, and inputs from the operation and strategic levels.  Each context $\theta_i, i\in \bar{T}$ corresponds to a knowledge set $K_i, i\in \bar{T}$, which includes the rules of engagement, system constraints on security policy adjustments, permissible conjectures, and patterns for updating policies. The knowledge set affects the way how the defender conjectures the attacker and adapts the security policy at the tactical level.

FMs serve as pivotal tools across various functions including reinforcement learning, knowledge assimilation, formation of conjectures, and contextual representation. Moreover, they facilitate the coordination of multiple components in neurosymbolic learning. Figure \ref{fmgm} further illustrates the synergy between FMs and GMs for cyber deception. A sequence of tactical games $G_1, G_2, G_3$ unfold between an attacker and a defender. The defender leverages $F_1$ to conjecture the attacker's behavior and utilizes $F_2$ to adapt security measures based on acquired observations. This conjecture and subsequent updates stem from the defender's knowledge, which is symbolically represented using   $F_4$.

Driving the conjectural learning process is $F_5$, coordinating the evolution of conjectures and policy updates. The tactical game unfolds iteratively, driven by operational-level objectives. Environmental shifts and operational goals require further adaptations in the tactical gameplay.  Contextual parameters denoted by $\theta$ capture these changes. $F_3$ processes diverse data sources to identify contextual changes, while $F_6$ bridges tactical gameplay with contextual shifts. Each context $\theta$ has an associated knowledge set $K$, which dynamically updates in response to contextual changes.  FM $F_7$ provides a set of rules for policy updates and conjecture formation under a context. FM $F_4$ is used to figure out the set of relevant knowledge which $F_7$ can choose from. $F_4$ can also create new rules for new contexts transferrable from other contexts or learned from datasets. 

The multi-scale perspective of neurosymbolic learning aligns with the multi-level representation of security games. Contextual and knowledge updates at the broader time scale correlate with operational-level games, which adapt to changes in mission and environment. The symbiotic interaction between GM and FM at this level fosters operational-level resilience. Conversely, security updates and attacker conjectures at the finer time scale correspond to tactical-level games, facilitating comprehensive adversarial reasoning and tactical responses. The symbiotic relationship between GM and FM at this level fosters tactical-level agility. Integrating these elements confers a decision-dominant advantage, providing both agility and resilience for the mission.

\begin{figure}
\begin{center}
    \includegraphics[scale=0.7]{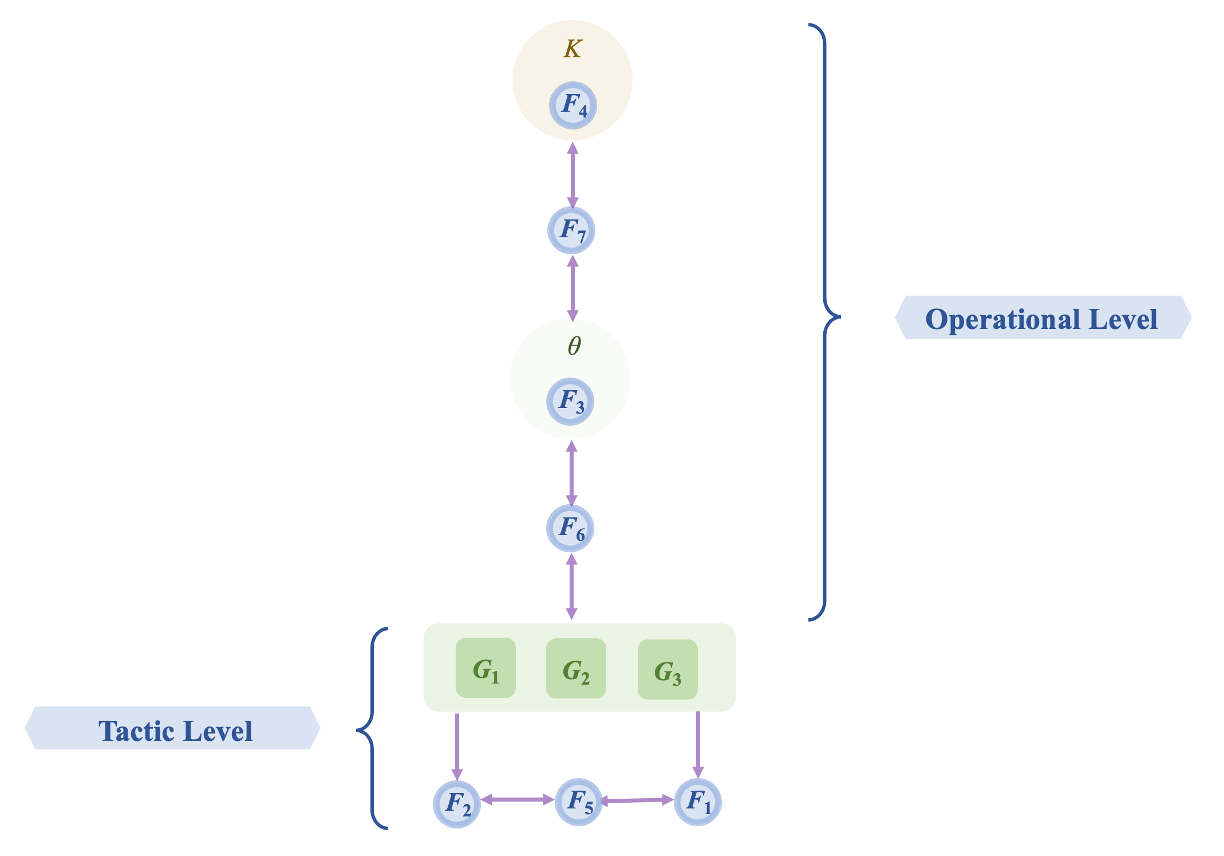}
    \caption{Interconnections between FMs and GMs for cyber deception: The tactical games are played between an attacker and a defender. A defender uses $F_1$ to conjecture the attacker's behavior and uses $F_2$ to update the security knowledge. The conjecture and the update are based on the defender's knowledge. The knowledge is symbolically represented and captured using $F_4$. The conjectured learning is driven by $F_5$, which coordinates the conjecture update and the policy update. The tactical game is sequentially played over time to achieve its operational-level objective. The changes in the environment and the operational-level objective can lead to the adaptation of tactical-level games. This change is represented by the contextual parameter $\theta$, which requires $F_5$ to process heterogeneous data sources to identify the context and $F_6$ to coordinate between the tactical level games and the contextual changes. Each context has an associated knowledge set $K$, which updates itself when the context changes.  FM $F_7$ provides a set of rules for policy updates and conjecture formation under a context. FM $F_4$ is used to figure out the set of relevant knowledge which $F_7$ can choose from. $F_4$ can also create new rules for new contexts transferrable from other contexts or learned from datasets. }
\end{center}
\end{figure}\label{fmgm}


\subsection{Emerging Challenges}

The practical implementation of security games encounters additional system constraints specific to the domain, hardware, and user requirements. Solutions must be tailored to these constraints, considering factors such as response time, ethical defense actions, temporal sequences, and hardware complexity for online learning. To address these constraints, FMs can be imposed during the training or design phase, for instance, hardware-related ones.

For certain constraints, especially logical ones, FMs can be employed to create discriminators capable of vetoing recommendations that do not align with a learned criterion specific to the application setting. This transition from a general-purpose security game to a specific-purpose one is achieved through transfer learning using an appropriate architecture of hyper-FMs.

Despite these promising directions, challenges persist. The foremost challenge is the scarcity of security data, creating the need for the amalgamation of domain knowledge, FMs, and agent simulators to fulfill design goals. The incorporation of system vulnerabilities, application workflows, and common attacker behaviors into the design process becomes crucial.

Another significant challenge revolves around ensuring high confidence in analytics and design. Given that security applications often pertain to mission-critical scenarios, the trustworthiness and reliability of utilizing FMs become paramount concerns. Strategies to address this challenge include the development of adaptive or tunable FMs and the establishment of feedback mechanisms, both contributing to achieving improved reliability in security analytics and design.

Another practical challenge arises from the inference time current FMs consume to output the next token. In the presence of advanced adversarial attacks, and in particular, cyber threats, real-time inference is the prerequisite of a rapid defense response to counteract attackers’ swift moves, such as lateral movement within the system. Accelerating the FM’s inference with long contexts is imperative for predictive and prescriptive security analytics.

\section{Conclusions}

Cyber deception operations are increasingly prevalent in today's cyber warfare, aiming to thwart and deter resourceful and intelligent attackers from targeted assets. They utilize decoys, honeypots, and misinformation to manipulate adversaries' behaviors and decision-making processes. This chapter investigates the relationship between game models (GMs) and foundation models (FMs) for cyber deception. On one hand, GMs symbolically encode nominal knowledge of interactions between attackers and defenders. The game-theoretic representation, in strategic, extensive, or learning forms, encapsulates knowledge and experience of attack models and potential defense outcomes. On the other hand, FMs are baseline machine learning models providing powerful tools to process and extract information from heterogeneous data, including unstructured text data such as risk reports, regulatory documents, and incident reports. They enable predictive decision-making by simulating and generating known or unknown scenarios based on historical and current data, facilitating proactive and adaptive responses to attackers online. GMs and FMs can be symbiotically integrated to achieve multiple cyber defense functions, leading to guardware that transforms network security. These building blocks can be integrated through diverse architectures for descriptive, prescriptive, and predictive analytics for applications at tactical, operational, and strategic levels of the mission. This chapter proposes a multi-agent neurosymbolic reinforcement learning paradigm integrating GMs and FMs into one learning-based framework. It advocates for the capabilities of predictive intelligence, symbolic reasoning, and knowledge update in defensive cyber deception. Despite challenges in FMs and their integration with GMs, this chapter provides a promising path toward cyber resilience and decision dominance in cyber warfare.

\bibliographystyle{spbasic-unsort}

\bibliography{reference.bib} 

\begin{thebibliography}{115}
\providecommand{\natexlab}[1]{#1}
\providecommand{\url}[1]{{#1}}
\providecommand{\urlprefix}{URL }
\expandafter\ifx\csname urlstyle\endcsname\relax
  \providecommand{\doi}[1]{DOI~\discretionary{}{}{}#1}\else
  \providecommand{\doi}{DOI~\discretionary{}{}{}\begingroup \urlstyle{rm}\Url}\fi
\providecommand{\eprint}[2][]{\url{#2}}

\bibitem[{Jajodia et~al.(2016)Jajodia, Subrahmanian, Swarup, and Wang}]{jajodia2016cyber}
Jajodia S, Subrahmanian V, Swarup V, Wang C (2016) Cyber deception. Cham, Switzerland: Springer

\bibitem[{Al-Shaer et~al.(2019)Al-Shaer, Wei, Kevin, and Wang}]{al2019autonomous}
Al-Shaer E, Wei J, Kevin W, Wang C (2019) Autonomous cyber deception. Springer

\bibitem[{Li et~al.(2022)Li, Zhao, and Zhu}]{tao_info}
Li T, Zhao Y, Zhu Q (2022) {The role of information structures in game-theoretic multi-agent learning}. Annual Reviews in Control 53:296--314, \doi{10.1016/j.arcontrol.2022.03.003}

\bibitem[{Zhu et~al.(2010)Zhu, Li, Han, and Ba{\c{s}}ar}]{zhu2010stochastic}
Zhu Q, Li H, Han Z, Ba{\c{s}}ar T (2010) A stochastic game model for jamming in multi-channel cognitive radio systems. In: 2010 IEEE International Conference on Communications, IEEE, pp 1--6

\bibitem[{Zhu et~al.(2011)Zhu, Saad, Han, Poor, and Ba{\c{s}}ar}]{zhu2011eavesdropping}
Zhu Q, Saad W, Han Z, Poor HV, Ba{\c{s}}ar T (2011) Eavesdropping and jamming in next-generation wireless networks: A game-theoretic approach. In: 2011-MILCOM 2011 Military Communications Conference, IEEE, pp 119--124

\bibitem[{Xu and Zhu(2017)}]{xu2017game}
Xu Z, Zhu Q (2017) A game-theoretic approach to secure control of communication-based train control systems under jamming attacks. In: Proceedings of the 1st International Workshop on Safe Control of Connected and Autonomous Vehicles, pp 27--34

\bibitem[{Nugraha et~al.(2019)Nugraha, Hayakawa, Cetinkaya, Ishii, and Zhu}]{nugraha2019subgame}
Nugraha Y, Hayakawa T, Cetinkaya A, Ishii H, Zhu Q (2019) Subgame perfect equilibrium analysis for jamming attacks on resilient graphs. In: 2019 American Control Conference (ACC), IEEE, pp 2060--2065

\bibitem[{Ge and Zhu(2023)}]{ge2023gazeta}
Ge Y, Zhu Q (2023) Gazeta: Game-theoretic zero-trust authentication for defense against lateral movement in 5g iot networks. IEEE Transactions on Information Forensics and Security

\bibitem[{Rass et~al.(2020)Rass, Schauer, K{\"o}nig, Zhu, Rass, Schauer, K{\"o}nig, and Zhu}]{rass2020cryptographic}
Rass S, Schauer S, K{\"o}nig S, Zhu Q, Rass S, Schauer S, K{\"o}nig S, Zhu Q (2020) Cryptographic games. Cyber-Security in Critical Infrastructures: A Game-Theoretic Approach pp 223--247

\bibitem[{Ge and Zhu(2022)}]{ge2022mufaza}
Ge Y, Zhu Q (2022) Mufaza: Multi-source fast and autonomous zero-trust authentication for 5g networks. In: MILCOM 2022-2022 IEEE Military Communications Conference (MILCOM), IEEE, pp 571--576

\bibitem[{Gupta et~al.(2022)Gupta, Kumar, Shekhar, Sharma, Patel, Jain, Dhaou, and Iwendi}]{gupta2022game}
Gupta M, Kumar R, Shekhar S, Sharma B, Patel RB, Jain S, Dhaou IB, Iwendi C (2022) Game theory-based authentication framework to secure internet of vehicles with blockchain. Sensors 22(14):5119

\bibitem[{Sar{\i}ta{\c{s}} et~al.(2019)Sar{\i}ta{\c{s}}, Shereen, Sandberg, and D{\'a}n}]{saritacs2019adversarial}
Sar{\i}ta{\c{s}} S, Shereen E, Sandberg H, D{\'a}n G (2019) Adversarial attacks on continuous authentication security: A dynamic game approach. In: International Conference on Decision and Game Theory for Security, Springer, pp 439--458

\bibitem[{Zhu et~al.(2012{\natexlab{a}})Zhu, Clark, Poovendran, and Ba{\c{s}}ar}]{zhu2012deceptive}
Zhu Q, Clark A, Poovendran R, Ba{\c{s}}ar T (2012{\natexlab{a}}) Deceptive routing games. In: 2012 IEEE 51st IEEE Conference on Decision and Control (CDC), IEEE, pp 2704--2711

\bibitem[{Zhu et~al.(2012{\natexlab{b}})Zhu, Yuan, Song, Han, and Basar}]{zhu2012interference}
Zhu Q, Yuan Z, Song JB, Han Z, Basar T (2012{\natexlab{b}}) Interference aware routing game for cognitive radio multi-hop networks. IEEE Journal on Selected Areas in Communications 30(10):2006--2015

\bibitem[{Clark et~al.(2012)Clark, Zhu, Poovendran, and Ba{\c{s}}ar}]{clark2012deceptive}
Clark A, Zhu Q, Poovendran R, Ba{\c{s}}ar T (2012) Deceptive routing in relay networks. In: Decision and Game Theory for Security: Third International Conference, GameSec 2012, Budapest, Hungary, November 5-6, 2012. Proceedings 3, Springer, pp 171--185

\bibitem[{Rass et~al.(2017)Rass, Alshawish, Abid, Schauer, Zhu, and De~Meer}]{rass2017physical}
Rass S, Alshawish A, Abid MA, Schauer S, Zhu Q, De~Meer H (2017) Physical intrusion games—optimizing surveillance by simulation and game theory. IEEE Access 5:8394--8407

\bibitem[{Hu and Zhu(2022)}]{hu2022evasion}
Hu Y, Zhu Q (2022) Evasion-aware neyman-pearson detectors: A game-theoretic approach. In: 2022 IEEE 61st Conference on Decision and Control (CDC), IEEE, pp 6111--6117

\bibitem[{Zhu and Ba{\c{s}}ar(2009)}]{zhu2009dynamic}
Zhu Q, Ba{\c{s}}ar T (2009) Dynamic policy-based ids configuration. In: Proceedings of the 48h IEEE Conference on Decision and Control (CDC) held jointly with 2009 28th Chinese Control Conference, IEEE, pp 8600--8605

\bibitem[{Zhu et~al.(2010)Zhu, Tembine, and Ba{\c{s}}ar}]{zhu2010network}
Zhu Q, Tembine H, Ba{\c{s}}ar T (2010) Network security configurations: A nonzero-sum stochastic game approach. In: Proceedings of the 2010 American control conference, IEEE, pp 1059--1064

\bibitem[{Chen and Zhu(2019)}]{chen2019game}
Chen J, Zhu Q (2019) A game-and decision-theoretic approach to resilient interdependent network analysis and design. Springer

\bibitem[{Huang et~al.(2018)Huang, Chen, and Zhu}]{huang2018factored}
Huang L, Chen J, Zhu Q (2018) Factored markov game theory for secure interdependent infrastructure networks. Game Theory for Security and Risk Management: From Theory to Practice pp 99--126

\bibitem[{Chen et~al.(2019)Chen, Touati, and Zhu}]{chen2019dynamic}
Chen J, Touati C, Zhu Q (2019) A dynamic game approach to strategic design of secure and resilient infrastructure network. IEEE Transactions on Information Forensics and security 15:462--474

\bibitem[{Chen et~al.(2017)Chen, Touati, and Zhu}]{chen2017dynamic}
Chen J, Touati C, Zhu Q (2017) A dynamic game analysis and design of infrastructure network protection and recovery: 125. ACM SIGMETRICS Performance Evaluation Review 45(2):128

\bibitem[{Huang et~al.(2017)Huang, Chen, and Zhu}]{huang2017large}
Huang L, Chen J, Zhu Q (2017) A large-scale markov game approach to dynamic protection of interdependent infrastructure networks. In: International Conference on Decision and Game Theory for Security, Springer, pp 357--376

\bibitem[{Huang and Zhu(2021)}]{huang2021duplicity}
Huang L, Zhu Q (2021) Duplicity games for deception design with an application to insider threat mitigation. IEEE Transactions on Information Forensics and Security 16:4843--4856

\bibitem[{Casey et~al.(2016)Casey, Morales, Wright, Zhu, and Mishra}]{casey2016compliance}
Casey W, Morales JA, Wright E, Zhu Q, Mishra B (2016) Compliance signaling games: toward modeling the deterrence of insider threats. Computational and Mathematical Organization Theory 22:318--349

\bibitem[{Casey et~al.(2015)Casey, Zhu, Morales, and Mishra}]{casey2015compliance}
Casey WA, Zhu Q, Morales JA, Mishra B (2015) Compliance control: Managed vulnerability surface in social-technological systems via signaling games. In: Proceedings of the 7th ACM CCS international workshop on managing insider security threats, pp 53--62

\bibitem[{Feng et~al.(2015)Feng, Zheng, Hu, Cansever, and Mohapatra}]{feng2015stealthy}
Feng X, Zheng Z, Hu P, Cansever D, Mohapatra P (2015) Stealthy attacks meets insider threats: A three-player game model. In: MILCOM 2015-2015 IEEE Military Communications Conference, IEEE, pp 25--30

\bibitem[{Liu et~al.(2008)Liu, Wang, and Camp}]{liu2008game}
Liu D, Wang X, Camp J (2008) Game-theoretic modeling and analysis of insider threats. International Journal of Critical Infrastructure Protection 1:75--80

\bibitem[{Rass and Schauer(2018)}]{rass2018game}
Rass S, Schauer S (2018) Game theory for security and risk management. Springer International Publishing doi 10:978--3

\bibitem[{Chen et~al.(2021)Chen, Zhu, and Ba{\c{s}}ar}]{chen2021dynamic}
Chen J, Zhu Q, Ba{\c{s}}ar T (2021) Dynamic contract design for systemic cyber risk management of interdependent enterprise networks. Dynamic Games and Applications 11:294--325

\bibitem[{Chen and Zhu(2018)}]{chen2018linear}
Chen J, Zhu Q (2018) A linear quadratic differential game approach to dynamic contract design for systemic cyber risk management under asymmetric information. In: 2018 56th Annual Allerton Conference on Communication, Control, and Computing (Allerton), IEEE, pp 575--582

\bibitem[{Zhang et~al.(2017)Zhang, Zhu, and Hayel}]{zhang2017bi}
Zhang R, Zhu Q, Hayel Y (2017) A bi-level game approach to attack-aware cyber insurance of computer networks. IEEE Journal on Selected Areas in Communications 35(3):779--794

\bibitem[{Zhang and Zhu(2019)}]{zhang2019mathtt}
Zhang R, Zhu Q (2019) {FlipIn}: A game-theoretic cyber insurance framework for incentive-compatible cyber risk management of internet of things. IEEE Transactions on Information Forensics and Security 15:2026--2041

\bibitem[{Schwartz and Sastry(2014)}]{schwartz2014cyber}
Schwartz GA, Sastry SS (2014) Cyber-insurance framework for large scale interdependent networks. In: Proceedings of the 3rd international conference on High confidence networked systems, pp 145--154

\bibitem[{Zhang and Zhu(2021)}]{zhang2021optimal}
Zhang R, Zhu Q (2021) Optimal cyber-insurance contract design for dynamic risk management and mitigation. IEEE Transactions on Computational Social Systems 9(4):1087--1100

\bibitem[{Huang and Zhu(2020)}]{huang2020dynamic}
Huang L, Zhu Q (2020) A dynamic games approach to proactive defense strategies against advanced persistent threats in cyber-physical systems. Computers \& Security 89:101660

\bibitem[{Chen and Zhu(2022)}]{chen2022system}
Chen J, Zhu Q (2022) A system-of-systems approach to strategic cyber-defense and robust switching control design for cyber-physical wind energy systems. In: Security and Resilience of Control Systems: Theory and Applications, Springer, pp 177--202

\bibitem[{Zhu and Xu(2020)}]{zhu2020cross}
Zhu Q, Xu Z (2020) Cross-layer design for secure and resilient cyber-physical systems. Springer

\bibitem[{Cavusoglu et~al.(2008)Cavusoglu, Raghunathan, and Yue}]{cavusoglu2008decision}
Cavusoglu H, Raghunathan S, Yue WT (2008) Decision-theoretic and game-theoretic approaches to it security investment. Journal of Management Information Systems 25(2):281--304

\bibitem[{Grossklags and Johnson(2009)}]{grossklags2009uncertainty}
Grossklags J, Johnson B (2009) Uncertainty in the weakest-link security game. In: 2009 International Conference on Game Theory for Networks, IEEE, pp 673--682

\bibitem[{Chen and Zhu(2018)}]{chen2018security}
Chen J, Zhu Q (2018) Security investment under cognitive constraints: A gestalt nash equilibrium approach. In: 2018 52nd Annual Conference on Information Sciences and Systems (CISS), IEEE, pp 1--6

\bibitem[{Chen and Zhu(2019)}]{chen2019interdependent}
Chen J, Zhu Q (2019) Interdependent strategic security risk management with bounded rationality in the internet of things. IEEE Transactions on Information Forensics and Security 14(11):2958--2971

\bibitem[{Harsanyi(1968)}]{harsanyi1968games}
Harsanyi JC (1968) Games with incomplete information played by “bayesian” players part ii. bayesian equilibrium points. Management science 14(5):320--334

\bibitem[{Harsanyi(1995)}]{harsanyi1995games}
Harsanyi JC (1995) Games with incomplete information. The American Economic Review 85(3):291--303

\bibitem[{Bennett(1980)}]{bennett1980hypergames}
Bennett PG (1980) Hypergames: developing a model of conflict. Futures 12(6):489--507

\bibitem[{Wang et~al.(1989)Wang, Hipel, and Fraser}]{wang1989solution}
Wang M, Hipel KW, Fraser NM (1989) Solution concepts in hypergames. Applied Mathematics and Computation 34(3):147--171

\bibitem[{La et~al.(2016{\natexlab{a}})La, Quek, and Lee}]{la2016game}
La QD, Quek TQ, Lee J (2016{\natexlab{a}}) A game theoretic model for enabling honeypots in iot networks. In: 2016 IEEE International Conference on Communications (ICC), IEEE, pp 1--6

\bibitem[{La et~al.(2016{\natexlab{b}})La, Quek, Lee, Jin, and Zhu}]{la2016deceptive}
La QD, Quek TQ, Lee J, Jin S, Zhu H (2016{\natexlab{b}}) Deceptive attack and defense game in honeypot-enabled networks for the internet of things. IEEE Internet of Things Journal 3(6):1025--1035

\bibitem[{Boumkheld et~al.(2019)Boumkheld, Panda, Rass, and Panaousis}]{boumkheld2019honeypot}
Boumkheld N, Panda S, Rass S, Panaousis E (2019) Honeypot type selection games for smart grid networks. In: Decision and Game Theory for Security: 10th International Conference, GameSec 2019, Stockholm, Sweden, October 30--November 1, 2019, Proceedings 10, Springer, pp 85--96

\bibitem[{Huang and Zhu(2019)}]{huang2019dynamic}
Huang L, Zhu Q (2019) Dynamic bayesian games for adversarial and defensive cyber deception. Autonomous Cyber Deception: Reasoning, Adaptive Planning, and Evaluation of HoneyThings pp 75--97

\bibitem[{Caballero et~al.(2024)Caballero, Cooley, Banks, and Jenkins}]{caballero2024behavioral}
Caballero W, Cooley J, Banks D, Jenkins P (2024) A behavioral approach to repeated bayesian security games. The Annals of Applied Statistics 18(1):199--223

\bibitem[{Zhang and Zhu(2023)}]{zhang2023stochastic}
Zhang T, Zhu Q (2023) Stochastic game with interactive information acquisition: A fixed-point alignment principle. In: 2023 59th Annual Allerton Conference on Communication, Control, and Computing (Allerton), IEEE, pp 1--8

\bibitem[{Zhang and Zhu(2022)}]{zhang2022forward}
Zhang T, Zhu Q (2022) Forward-looking dynamic persuasion for pipeline stochastic bayesian game: A fixed-point alignment principle. arXiv preprint arXiv:220309725

\bibitem[{Zhang and Zhu(2021)}]{zhang2021equilibrium}
Zhang T, Zhu Q (2021) On the equilibrium elicitation of markov games through information design. arXiv preprint arXiv:210207152

\bibitem[{Li and Zhu(2023)}]{tao23pot}
Li T, Zhu Q (2023) {On the Price of Transparency: A Comparison Between Overt Persuasion and Covert Signaling}. 2023 62nd IEEE Conference on Decision and Control (CDC) 00:4267--4272, \doi{10.1109/cdc49753.2023.10383897}, \eprint{2304.00096}

\bibitem[{Li et~al.(2022)Li, Peng, Zhu, and Baar}]{tao22confluence}
Li T, Peng G, Zhu Q, Baar T (2022) {The Confluence of Networks, Games, and Learning a Game-Theoretic Framework for Multiagent Decision Making Over Networks}. IEEE Control Systems 42(4):35--67, \doi{10.1109/mcs.2022.3171478}

\bibitem[{Liu et~al.(2023)Liu, Li, and Zhu}]{shutian23erm}
Liu S, Li T, Zhu Q (2023) {Game-Theoretic Distributed Empirical Risk Minimization With Strategic Network Design}. IEEE Transactions on Signal and Information Processing over Networks 9:542--556, \doi{10.1109/tsipn.2023.3306106}

\bibitem[{Pan et~al.(2023)Pan, Li, and Zhu}]{pan-tao22noneq}
Pan Y, Li T, Zhu Q (2023) {On the Resilience of Traffic Networks under Non-Equilibrium Learning}. 2023 American Control Conference (ACC) 00:3484--3489, \doi{10.23919/acc55779.2023.10156139}

\bibitem[{Hammar et~al.(2024)Hammar, Li, Stadler, and Zhu}]{li2024automated}
Hammar K, Li T, Stadler R, Zhu Q (2024) Automated security response through online learning with adaptive conjectures. arXiv preprint arXiv:240212499

\bibitem[{Pawlick et~al.(2019)Pawlick, Colbert, and Zhu}]{pawlick2019game}
Pawlick J, Colbert E, Zhu Q (2019) A game-theoretic taxonomy and survey of defensive deception for cybersecurity and privacy. ACM Computing Surveys (CSUR) 52(4):1--28

\bibitem[{Pawlick et~al.(2018)Pawlick, Colbert, and Zhu}]{pawlick2018modeling}
Pawlick J, Colbert E, Zhu Q (2018) Modeling and analysis of leaky deception using signaling games with evidence. IEEE Transactions on Information Forensics and Security 14(7):1871--1886

\bibitem[{Hor{\'a}k et~al.(2017)Hor{\'a}k, Zhu, and Bo{\v{s}}ansk{\`y}}]{horak2017manipulating}
Hor{\'a}k K, Zhu Q, Bo{\v{s}}ansk{\`y} B (2017) Manipulating adversary’s belief: A dynamic game approach to deception by design for proactive network security. In: Decision and Game Theory for Security: 8th International Conference, GameSec 2017, Vienna, Austria, October 23-25, 2017, Proceedings, Springer, pp 273--294

\bibitem[{Li et~al.(2024)Li, Hammar, Stadler, and Zhu}]{li2024conjectural}
Li T, Hammar K, Stadler R, Zhu Q (2024) Conjectural online learning with first-order beliefs in asymmetric information stochastic games. arXiv preprint arXiv:240218781

\bibitem[{Anwar and Kamhoua(2020)}]{anwar2020game}
Anwar AH, Kamhoua C (2020) Game theory on attack graph for cyber deception. In: International Conference on Decision and Game Theory for Security, Springer, pp 445--456

\bibitem[{Liu and Zhu(2023)}]{liu2023information}
Liu S, Zhu Q (2023) Information manipulation in partially observable markov decision processes. arXiv preprint arXiv:231207862

\bibitem[{Zhang and Zhu(2020)}]{zhang2020deceptive}
Zhang Z, Zhu Q (2020) Deceptive kernel function on observations of discrete pomdp. arXiv preprint arXiv:200805585

\bibitem[{Huang et~al.(2019)Huang, Kavitha, and Zhu}]{huang2019continuous}
Huang Y, Kavitha V, Zhu Q (2019) Continuous-time markov decision processes with controlled observations. In: 2019 57th Annual Allerton Conference on Communication, Control, and Computing (Allerton), IEEE, pp 32--39

\bibitem[{Huang and Zhu(2021)}]{huang2021pursuit}
Huang Y, Zhu Q (2021) A pursuit-evasion differential game with strategic information acquisition. arXiv preprint arXiv:210205469

\bibitem[{Huang et~al.(2021)Huang, Chen, and Zhu}]{huang2021defending}
Huang Y, Chen J, Zhu Q (2021) Defending an asset with partial information and selected observations: A differential game framework. In: 2021 60th IEEE Conference on Decision and Control (CDC), IEEE, pp 2366--2373

\bibitem[{Roberson(2006)}]{roberson2006colonel}
Roberson B (2006) The colonel blotto game. Economic Theory 29(1):1--24

\bibitem[{Hart(2008)}]{hart2008discrete}
Hart S (2008) Discrete colonel blotto and general lotto games. International Journal of Game Theory 36(3-4):441--460

\bibitem[{Golman and Page(2009)}]{golman2009general}
Golman R, Page SE (2009) General blotto: games of allocative strategic mismatch. Public Choice 138:279--299

\bibitem[{Van~Dijk et~al.(2013)Van~Dijk, Juels, Oprea, and Rivest}]{van2013flipit}
Van~Dijk M, Juels A, Oprea A, Rivest RL (2013) Flipit: The game of “stealthy takeover”. Journal of Cryptology 26:655--713

\bibitem[{Laszka et~al.(2014)Laszka, Horvath, Felegyhazi, and Butty{\'a}n}]{laszka2014flipthem}
Laszka A, Horvath G, Felegyhazi M, Butty{\'a}n L (2014) Flipthem: Modeling targeted attacks with flipit for multiple resources. In: Decision and Game Theory for Security: 5th International Conference, GameSec 2014, Los Angeles, CA, USA, November 6-7, 2014. Proceedings 5, Springer, pp 175--194

\bibitem[{Bowers et~al.(2012)Bowers, Van~Dijk, Griffin, Juels, Oprea, Rivest, and Triandopoulos}]{bowers2012defending}
Bowers KD, Van~Dijk M, Griffin R, Juels A, Oprea A, Rivest RL, Triandopoulos N (2012) Defending against the unknown enemy: Applying flipit to system security. In: International Conference on Decision and Game Theory for Security, Springer, pp 248--263

\bibitem[{Chen et~al.(2020)Chen, Touati, and Zhu}]{8673619}
Chen J, Touati C, Zhu Q (2020) Optimal secure two-layer iot network design. IEEE Transactions on Control of Network Systems 7(1):398--409, \doi{10.1109/TCNS.2019.2906893}

\bibitem[{Zhu and Rass(2018)}]{zhu2018multi}
Zhu Q, Rass S (2018) On multi-phase and multi-stage game-theoretic modeling of advanced persistent threats. IEEE Access 6:13958--13971

\bibitem[{Manshaei et~al.(2013)Manshaei, Zhu, Alpcan, Bac{\c{s}}ar, and Hubaux}]{manshaei2013game}
Manshaei MH, Zhu Q, Alpcan T, Bac{\c{s}}ar T, Hubaux JP (2013) Game theory meets network security and privacy. ACM Computing Surveys (CSUR) 45(3):1--39

\bibitem[{Esponda and Pouzo(2016)}]{esponda2016berk}
Esponda I, Pouzo D (2016) Berk--nash equilibrium: A framework for modeling agents with misspecified models. Econometrica 84(3):1093--1130

\bibitem[{Nash(1953)}]{nash1953two}
Nash J (1953) Two-person cooperative games. Econometrica: Journal of the Econometric Society pp 128--140

\bibitem[{Li et~al.(2024)Li, Bian, Lei, Zuo, Yang, Zhu, Li, Chen, and Ozbay}]{tao2024dima}
Li T, Bian Z, Lei H, Zuo F, Yang YT, Zhu Q, Li Z, Chen Z, Ozbay K (2024) Digital twin-based driver risk-aware intelligent mobility analytics for urban transportation management. arXiv preprint arXiv:240715025

\bibitem[{Peng et~al.(2021)Peng, Li, Liu, Chen, and Zhu}]{peng21local}
Peng G, Li T, Liu S, Chen J, Zhu Q (2021) Locally-aware constrained games on networks. In: 2021 American Control Conference (ACC), pp 4606--4611, \doi{10.23919/ACC50511.2021.9482895}

\bibitem[{Yin et~al.(2024)Yin, Li, Lei, Hu, Rangan, and Zhu}]{yin24pirl}
Yin M, Li T, Lei H, Hu Y, Rangan S, Zhu Q (2024) Zero-shot wireless indoor navigation through physics-informed reinforcement learning. In: 2024 IEEE International Conference on Robotics and Automation (ICRA), pp 5111--5118, \doi{10.1109/ICRA57147.2024.10611229}

\bibitem[{OpenAI(2023)}]{openai23}
OpenAI (2023) {GPT-4} technical report. \urlprefix\url{https://arxiv.org/abs/2303.08774}

\bibitem[{Ramesh et~al.(2021)Ramesh, Pavlov, Goh, Gray, Voss, Radford, Chen, and Sutskever}]{ramesh21dalle}
Ramesh A, Pavlov M, Goh G, Gray S, Voss C, Radford A, Chen M, Sutskever I (2021) Zero-shot text-to-image generation. In: Meila M, Zhang T (eds) Proceedings of the 38th International Conference on Machine Learning, PMLR, Proceedings of Machine Learning Research, vol 139, pp 8821--8831, \urlprefix\url{https://proceedings.mlr.press/v139/ramesh21a.html}

\bibitem[{Copet et~al.(2024)Copet, Kreuk, Gat, Remez, Kant, Synnaeve, Adi, and D{\'e}fossez}]{copet2024simple}
Copet J, Kreuk F, Gat I, Remez T, Kant D, Synnaeve G, Adi Y, D{\'e}fossez A (2024) Simple and controllable music generation. Advances in Neural Information Processing Systems 36

\bibitem[{Zitkovich et~al.(2023)Zitkovich, Yu, Xu, Xu, Xiao, Xia, Wu, Wohlhart, Welker, Wahid, Vuong, Vanhoucke, Tran, Soricut, Singh, Singh, Sermanet, Sanketi, Salazar, Ryoo, Reymann, Rao, Pertsch, Mordatch, Michalewski, Lu, Levine, Lee, Lee, Leal, Kuang, Kalashnikov, Julian, Joshi, Irpan, Ichter, Hsu, Herzog, Hausman, Gopalakrishnan, Fu, Florence, Finn, Dubey, Driess, Ding, Choromanski, Chen, Chebotar, Carbajal, Brown, Brohan, Arenas, and Han}]{zitkovich23rt2}
Zitkovich B, Yu T, Xu S, Xu P, Xiao T, Xia F, Wu J, Wohlhart P, Welker S, Wahid A, Vuong Q, Vanhoucke V, Tran H, Soricut R, Singh A, Singh J, Sermanet P, Sanketi PR, Salazar G, Ryoo MS, Reymann K, Rao K, Pertsch K, Mordatch I, Michalewski H, Lu Y, Levine S, Lee L, Lee TWE, Leal I, Kuang Y, Kalashnikov D, Julian R, Joshi NJ, Irpan A, Ichter B, Hsu J, Herzog A, Hausman K, Gopalakrishnan K, Fu C, Florence P, Finn C, Dubey KA, Driess D, Ding T, Choromanski KM, Chen X, Chebotar Y, Carbajal J, Brown N, Brohan A, Arenas MG, Han K (2023) Rt-2: Vision-language-action models transfer web knowledge to robotic control. In: Tan J, Toussaint M, Darvish K (eds) Proceedings of The 7th Conference on Robot Learning, PMLR, Proceedings of Machine Learning Research, vol 229, pp 2165--2183, \urlprefix\url{https://proceedings.mlr.press/v229/zitkovich23a.html}

\bibitem[{Vaswani et~al.(2017)Vaswani, Shazeer, Parmar, Uszkoreit, Jones, Gomez, Kaiser, and Polosukhin}]{vaswani17transformer}
Vaswani A, Shazeer N, Parmar N, Uszkoreit J, Jones L, Gomez AN, Kaiser L, Polosukhin I (2017) {Attention is All you Need}. In: Advances in Neural Information Processing Systems, Curran Associates, Inc., NeuriPS, vol~30, \doi{10.48550/arxiv.1706.03762}, \urlprefix\url{https://proceedings.neurips.cc/paper\_files/paper/2017/file/3f5ee243547dee91fbd053c1c4a845aa-Paper.pdf}

\bibitem[{Ranade et~al.(2021)Ranade, Piplai, Joshi, and Finin}]{cybert21}
Ranade P, Piplai A, Joshi A, Finin T (2021) Cybert: Contextualized embeddings for the cybersecurity domain. In: 2021 IEEE International Conference on Big Data (Big Data), pp 3334--3342, \doi{10.1109/BigData52589.2021.9671824}

\bibitem[{Das et~al.(2022)Das, Halappanavar, Tumeo, Serra, Pothen, and Al-Shaer}]{cve-cwe}
Das SS, Halappanavar M, Tumeo A, Serra E, Pothen A, Al-Shaer E (2022) Vwc-bert: Scaling vulnerability–weakness–exploit mapping on modern ai accelerators. In: 2022 IEEE International Conference on Big Data (Big Data), pp 1224--1229, \doi{10.1109/BigData55660.2022.10020622}

\bibitem[{Grigorescu et~al.(2022)Grigorescu, Nica, Dascalu, and Rughinis}]{cve2attck}
Grigorescu O, Nica A, Dascalu M, Rughinis R (2022) Cve2att\&ck: Bert-based mapping of cves to mitre att\&ck techniques. Algorithms 15(9), \doi{10.3390/a15090314}, \urlprefix\url{https://www.mdpi.com/1999-4893/15/9/314}

\bibitem[{Bannon et~al.(2020)Bannon, Windsor, Song, and Li}]{tao2020causality}
Bannon J, Windsor B, Song W, Li T (2020) Causality and batch reinforcement learning: Complementary approaches to planning in unknown domains. arXiv preprint arXiv:200602579

\bibitem[{Li et~al.(2023)Li, Lei, and Zhu}]{tao23cola}
Li T, Lei H, Zhu Q (2023) {Self-Adaptive Driving in Nonstationary Environments through Conjectural Online Lookahead Adaptation}. 2023 IEEE International Conference on Robotics and Automation (ICRA) 00:7205--7211, \doi{10.1109/icra48891.2023.10161368}, \eprint{2210.03209}

\bibitem[{Janner et~al.(2021)Janner, Li, and Levine}]{janner2021offline}
Janner M, Li Q, Levine S (2021) Offline reinforcement learning as one big sequence modeling problem. Advances in neural information processing systems 34:1273--1286

\bibitem[{Li et~al.(2024)Li, Bian, Lei, Zuo, Yang, Zhu, Li, and Ozbay}]{tao24picol}
Li T, Bian Z, Lei H, Zuo F, Yang YT, Zhu Q, Li Z, Ozbay K (2024) {Multi-level traffic-responsive tilt camera surveillance through predictive correlated online learning}. Transportation Research Part C: Emerging Technologies 167:104804, \doi{10.1016/j.trc.2024.104804}

\bibitem[{Li et~al.(2023)Li, Guevara, Xie, and Zhu}]{tao23sce}
Li T, Guevara J, Xie X, Zhu Q (2023) {Self-Confirming Transformer for Locally Consistent Online Adaptation in Multi-Agent Reinforcement Learning}. arXiv \doi{10.48550/arxiv.2310.04579}, \eprint{2310.04579}

\bibitem[{Pan et~al.(2023)Pan, Li, and Zhu}]{pan-tao23delay}
Pan Y, Li T, Zhu Q (2023) {Is Stochastic Mirror Descent Vulnerable to Adversarial Delay Attacks? A Traffic Assignment Resilience Study}. 2023 62nd IEEE Conference on Decision and Control (CDC) 00:8328--8333, \doi{10.1109/cdc49753.2023.10384003}, \eprint{2304.01161}

\bibitem[{Pan et~al.(2024)Pan, Li, and Zhu}]{pan2024variational}
Pan Y, Li T, Zhu Q (2024) On the variational interpretation of mirror play in monotone games. arXiv preprint arXiv:240315636

\bibitem[{Li et~al.(2024)Li, Pan, and Zhu}]{li23dd-ztd}
Li T, Pan Y, Zhu Q (2024) {Decision-Dominant Strategic Defense Against Lateral Movement for 5G Zero-Trust Multi-Domain Networks}. In: Chen, Wu Y, Yu J, Wang P, Xiaogang (eds) Network Security Empowered by Artificial Intelligence, Springer Nature Switzerland, Cham, pp 25--76, \doi{10.1007/978-3-031-53510-9\_2}, \urlprefix\url{https://doi.org/10.1007/978-3-031-53510-9\_2}

\bibitem[{Li et~al.(2021)Li, Peng, and Zhu}]{Tao_blackwell}
Li T, Peng G, Zhu Q (2021) {Blackwell Online Learning for Markov Decision Processes}. 2021 55th Annual Conference on Information Sciences and Systems (CISS) 00:1--6, \doi{10.1109/ciss50987.2021.9400319}

\bibitem[{Ge et~al.(2023)Ge, Li, and Zhu}]{tao23ztd}
Ge Y, Li T, Zhu Q (2023) {Scenario-Agnostic Zero-Trust Defense with Explainable Threshold Policy: A Meta-Learning Approach}. IEEE INFOCOM 2023 - IEEE Conference on Computer Communications Workshops (INFOCOM WKSHPS) 00:1--6, \doi{10.1109/infocomwkshps57453.2023.10225816}, \eprint{2303.03349}

\bibitem[{Pan et~al.(2023)Pan, Li, Li, Xu, Zheng, and Zhu}]{pan-tao23meta-sg}
Pan Y, Li T, Li H, Xu T, Zheng Z, Zhu Q (2023) {A First Order Meta Stackelberg Method for Robust Federated Learning}. In: Adversarial Machine Learning Frontiers Workshop at 40th International Conference on Machine Learning, \doi{10.48550/arxiv.2306.13800}

\bibitem[{Michael et~al.(2020)Michael, Eilon, and Shmuel}]{solan_game}
Michael M, Eilon S, Shmuel Z (2020) {Game Theory}. Cambridge University Press, Cambridge, \doi{10.1017/cbo9780511794216}, \urlprefix\url{https://www.cambridge.org/core/books/game-theory/B0C072F66E027614E46A5CAB26394C7D}

\bibitem[{Fudenberg and Levine(1993)}]{self-confirming}
Fudenberg D, Levine DK (1993) {Self-Confirming Equilibrium}. Econometrica 61(3):523, \doi{10.2307/2951716}

\bibitem[{Huang et~al.(2022)Huang, Jia, Balcetis, and Zhu}]{huang2022advert}
Huang L, Jia S, Balcetis E, Zhu Q (2022) Advert: an adaptive and data-driven attention enhancement mechanism for phishing prevention. IEEE Transactions on Information Forensics and Security 17:2585--2597

\bibitem[{Yang et~al.(2023)Yang, Li, and Zhu}]{yating-tao23information}
Yang YT, Li T, Zhu Q (2023) {Designing Policies for Truth: Combating Misinformation with Transparency and Information Design}. 2023 21st International Symposium on Modeling and Optimization in Mobile, Ad Hoc, and Wireless Networks (WiOpt) 00:127--134, \doi{10.23919/wiopt58741.2023.10349848}

\bibitem[{Kamhoua et~al.(2021)Kamhoua, Kiekintveld, Fang, and Zhu}]{kamhoua2021game}
Kamhoua CA, Kiekintveld CD, Fang F, Zhu Q (2021) Game theory and machine learning for cyber security. John Wiley \& Sons

\bibitem[{Chen et~al.(2021)Chen, Lu, Rajeswaran, Lee, Grover, Laskin, Abbeel, Srinivas, and Mordatch}]{chen21dt}
Chen L, Lu K, Rajeswaran A, Lee K, Grover A, Laskin M, Abbeel P, Srinivas A, Mordatch I (2021) Decision transformer: Reinforcement learning via sequence modeling. In: Ranzato M, Beygelzimer A, Dauphin Y, Liang P, Vaughan JW (eds) Advances in Neural Information Processing Systems, Curran Associates, Inc., vol~34, pp 15084--15097, \urlprefix\url{https://proceedings.neurips.cc/paper_files/paper/2021/file/7f489f642a0ddb10272b5c31057f0663-Paper.pdf}

\bibitem[{Fung et~al.(2011)Fung, Zhu, Boutaba, and Ba{\c{s}}ar}]{fung2011smurfen}
Fung C, Zhu Q, Boutaba R, Ba{\c{s}}ar T (2011) Smurfen: A system framework for rule sharing collaborative intrusion detection. In: 2011 7th International Conference on Network and Service Management, IEEE, pp 1--6

\bibitem[{Zhu et~al.(2011)Zhu, Fung, Boutaba, and Ba{\c{s}}ar}]{zhu2011game}
Zhu Q, Fung C, Boutaba R, Ba{\c{s}}ar T (2011) A game-theoretic approach to rule sharing mechanism in networked intrusion detection systems: Robustness, incentives and security. In: 2011 50th IEEE Conference on Decision and Control and European Control Conference, IEEE, pp 243--248

\bibitem[{Zhu et~al.(2012)Zhu, Fung, Boutaba, and Basar}]{zhu2012guidex}
Zhu Q, Fung C, Boutaba R, Basar T (2012) Guidex: A game-theoretic incentive-based mechanism for intrusion detection networks. IEEE Journal on Selected Areas in Communications 30(11):2220--2230

\bibitem[{West et~al.(2022)West, Bhagavatula, Hessel, Hwang, Jiang, Le~Bras, Lu, Welleck, and Choi}]{west2022symbolic}
West P, Bhagavatula C, Hessel J, Hwang J, Jiang L, Le~Bras R, Lu X, Welleck S, Choi Y (2022) Symbolic knowledge distillation: from general language models to commonsense models. In: Carpuat M, de~Marneffe MC, Meza~Ruiz IV (eds) Proceedings of the 2022 Conference of the North American Chapter of the Association for Computational Linguistics: Human Language Technologies, Association for Computational Linguistics, Seattle, United States, pp 4602--4625, \doi{10.18653/v1/2022.naacl-main.341}, \urlprefix\url{https://aclanthology.org/2022.naacl-main.341}

\bibitem[{Wu et~al.(2023)Wu, Li, Sun, and Lu}]{wu23llm-symbolic}
Wu X, Li YL, Sun J, Lu C (2023) Symbol-llm: Leverage language models for symbolic system in visual human activity reasoning. In: Oh A, Neumann T, Globerson A, Saenko K, Hardt M, Levine S (eds) Advances in Neural Information Processing Systems, Curran Associates, Inc., vol~36, pp 29680--29691, \urlprefix\url{https://proceedings.neurips.cc/paper_files/paper/2023/file/5edb57c05c81d04beb716ef1d542fe9e-Paper-Conference.pdf}

\bibitem[{Huang and Zhu(2019)}]{huang2019adaptive}
Huang L, Zhu Q (2019) Adaptive honeypot engagement through reinforcement learning of semi-markov decision processes. In: Decision and Game Theory for Security: 10th International Conference, GameSec 2019, Stockholm, Sweden, October 30--November 1, 2019, Proceedings 10, Springer, pp 196--216

\end{thebibliography}
\end{document}